\documentclass[a4paper]{article}
\usepackage{amscd,amsmath,amssymb}
\usepackage[usenames, dvipsnames]{color}

\newcommand{\hC}{\widehat{C}}
\newcommand{\ha}{\hat{\alpha}}
\newcommand{\hb}{\hat{\beta}}
\newcommand{\hg}{\hat{\gamma}}

\newcommand{\p}[1]{(\ref{#1})}

\newcommand{\ad}{{\sf ad}}
\newcommand{\aA}{{{\mathbb{A}\mathbb{C}}}}
\newcommand{\sS}{{{\mathbb{S}\mathbb{C}}}}
\newcommand{\rS}{{{\mathbb{R}\mathbb{C}}}}

\newcommand{\sI}{{{\mathbb{S}}}}

\newcommand{\kK}{{{\mathbb{K}}}}

\newcommand{\sla}{\mathfrak{sl}}
\newcommand{\soa}{\mathfrak{so}}
\newcommand{\spa}{\mathfrak{sp}}

\def\lb{\label}
\def\qed{\rule{5pt}{5pt}}

\newcommand{\be}{\begin{equation}}
\newcommand{\ee}{\end{equation}}
\newcommand{\bea}{\begin{eqnarray}}
\newcommand{\eea}{\end{eqnarray}}

\newcommand{\ba}{\begin{array}} \newcommand{\ea}{\end{array}}

\newcommand{\nn}{\nonumber}
\def\theequation{\arabic{section}.\arabic{equation}}

\DeclareMathOperator{\id}{\mathsf{id}}
\DeclareMathOperator{\proj}{P}
\DeclareMathOperator{\projt}{\widetilde{P}}
\DeclareMathOperator{\Tr}{Tr}

\newcommand{\dimg}{\dim\mkern-2mu\mathfrak{g}}

\begin{document}

\begin{flushright}
\today \\
\end{flushright}
\vspace{2.3cm}

\begin{center}
{\huge\bf Split Casimir operator for simple Lie algebras
in the cube of $\ad$-representation  and Vogel parameters}
\end{center}
\vspace{1cm}

\begin{center}
{\Large \bf  A.P. Isaev${}^{a,b}$, S.O. Krivonos${}^{a}$
and A.A. Provorov${}^{a,c}$}
\end{center}

\vspace{0.2cm}

\begin{center}
{${}^a$ \it
Bogoliubov  Laboratory of Theoretical Physics,\\
Joint Institute for Nuclear Research,
141980 Dubna, Russia}\vspace{0.1cm}

{${}^b$ \it Lomonosov Moscow State University, \\
 Physics Faculty, Russia}\vspace{0.1cm}
 
{${}^c$ \it Moscow Institute of Physics and Technology (National Research University), Dolgoprudny, Moscow Region, Russia}\vspace{0.3cm}

{\tt isaevap@theor.jinr.ru, krivonos@theor.jinr.ru,\\[0.2cm] aleksanderprovorov@gmail.com}
\end{center}
\vspace{3cm}

\begin{abstract}\noindent
  We constructed characteristic identities for the 3-split (polarized)
  Casimir operators of simple Lie algebras in the adjoint representations $\ad$ and deduced
a certain class of subrepresentations
  in $\ad^{\otimes 3}$. The
  projectors onto invariant subspaces for
these  subrepresentations were directly
 constructed from the characteristic identities
 for the 3-split Casimir operators. For all simple
 Lie algebras, 
 universal expressions for the traces of higher powers of the 3-split Casimir operators
 were found and
 dimensions of the subrepresentations in
 $\ad^{\otimes 3}$ were calculated.
All our formulas are in agreement with the universal description of (irreducible)
subrepresentations
in $\ad^{\otimes 3}$ for simple Lie algebras in terms of the Vogel parameters.
\end{abstract}

\newpage

\pagenumbering{arabic}
\setcounter{page}{2}
\section{Introduction}
It is known that a special invariant operator, the split (or polarized) Casimir operator
$\hC$ (see definition in Section 2), plays an important role in both describing  Lie algebras $\mathfrak{g}$ and studying of their representation theory. On the other hand,
the split Casimir operator $\hC$ is a building block (see, e.g., \cite{ChPr}, \cite{Ma} and references therein)
for constructing $\mathfrak{g}$-invariant solutions $r$ and $R$ of semiclassical and quantum Yang-Baxter
equations. Recall \cite{Dr} that $\mathfrak{g}$-invariant rational solutions of the Yang-Baxter equations allow one to define the Yangians $Y (\mathfrak{g})$ within the so-called $RTT$-realization. The Casimir operators are also found to be useful in computation of color factors for Feynman diagrams in non-Abelian gauge theories (see e.g., \cite{RSV}).

In this paper, we demonstrate the usefulness of the $\mathfrak{g}$-invariant split Casimir operator $\hC$ in the representation theory of Lie algebras. Namely, for all simple Lie algebras $\mathfrak{g}$, explicit formulas are found for invariant projectors onto irreducible representations that appear in the expansion of the tensor product $\ad^{\otimes 3}$ of three adjoint representations $\ad$. The idea to use $\mathfrak{g}$-invariant operators to construct such invariant projectors is not new. For example, invariant projectors acting in tensor representations of the $\mathfrak{sl}_n$ algebras are called Young symmetrizers and are constructed as images of special elements (idempotents) of the group algebra $\mathbb{C}[S_r]$ of the symmetric group $S_r$. The algebra $\mathbb{C}[S_r]$ centralizes the action of the $\mathfrak{sl}_n$ in the space of tensors of rank $r$ (in the spaces of tensor powers $T_f^{\otimes r}$ of the defining representation $T_f$). For the $\mathfrak{so}_n$ Lie algebras there exists an analogous statement: the action of those algebras in the representation $T^{\otimes r}_f$ is centralized by the Brauer algebra $B_r(n)$ (see e.g. \cite{Book2}). The use of the split Casimir operator, however, allows one to  decompose the tensor power of the representation into irreducible ones uniformly for all simple Lie algebras.

Our study of the split Casimir operator $\hC$ was also motivated by the works \cite{MSV, MkrV, MMM, MMM1} and the notion of the universal Lie algebra, which was introduced by P. Vogel in \cite{Fog} (see also \cite{Delig,Lan}; see also historical remarks in \cite{Cvit}, Sect.21.2). The universal Lie algebra was supposed to be a model of all complex simple Lie algebras $\mathfrak{g}$. For example, many quantities that characterize an algebra $\mathfrak{g}$ in different representations $T_\lambda$ (possibly reducible) entering into the decomposition $\ad^{\otimes k}=\sum_\lambda T_\lambda$ of $\ad^{\otimes k}$, where $k \ge 1$, can be expressed analytically as functions of  three Vogel parameters (see their definition in Section 2). These parameters take specific values for each of the complex simple Lie algebras $\mathfrak{g}$ (see e.g. \cite{MSV}, and Section 2 below). In particular, it was shown that using the Vogel parameters, one can express the dimensions of the representations $T_\lambda$ in the cases of $k = 2, 3$ \cite{Fog}, the dimensions of the representations $T_\lambda$ in the cases of $k=2,3,4$ for exceptional Lie algebras \cite{Cohen}, the dimension of an arbitrary representation $T_{\lambda'}$ with the highest weight $\lambda'= k \lambda_{\ad}$, where $\lambda_{\ad}$ is the highest root of $\mathfrak{g}$ \cite{Lan2}, the dimensions of the representations of $X_2$-series \cite{AvMkr} as well as values of the higher Casimir operators in the adjoint representation of $\mathfrak{g}$ \cite{MSV}. In \cite{MkrV2}, a universal formula for the volume of a connected and simply connected compact Lie group with Lie algebra $\mathfrak{g}$ was obtained. In \cite{MMM}, it was shown that the universal description of complex simple Lie algebras allows one to formulate several types of knot polynomials as a single function for all the simple Lie algebras simultaneously. The notion of Vogel's parameters allows one to formulate some new questions. For example, in \cite{Rub}, the condition for cancellation of singularities of the universal character formula for the adjoint representation of a simple Lie algebra leads to a set of Diophantine equations. Some of its solutions correspond to simple Lie algebras, but there are also some other points, the interpretation of which seems to be an intriguing problem.

In the original work \cite{Fog} of Vogel, the notion of the universal Lie algebra was obtained by means of highly nontrivial considerations stemming from the theory of knots. In \cite{Delig}, it was conjectured and in \cite{Cohen} it was proved that the representations $\ad^{\otimes 3}$ and $\ad^{\otimes 4}$ of exceptional Lie algebras admit a universal decomposition. In the present paper we rederive many of Vogel's results by taking an elementary approach of using the split Casimir operator $\hC$ to decompose $\ad^{\otimes 3}$ into subrepresentations for all simple Lie algebras. This shows that the use of the split Casimir operator provides a general simple and universal algorithm for expanding various Lie algebra representations as a direct sum of subrepresentations. Having used it to rederive the main results of \cite{Fog}, we plan on going further and applying this technique to find  a universal decomposition of the representation $\ad^{\otimes 4}$ for all simple Lie algebras.

The paper is organized as follows. In Section 2, we recall some of the basic notions of the Lie algebra theory, define Vogel's parameters, and introduce conventions that will be used throughout the work. In Sections 3, 4, and 5, we provide universal characteristic identities of $\hC$, universal dimensions of the corresponding subrepresentations, and derive the aforementioned identities in the special cases of simple Lie algebras in the symmetric, antisymmetric, and "hook" parts of $\hC$, respectively. In Section 6, we concisely summarize the results. Most of them we have obtained by means of a computer program, and in Appendix A we show that some of the formulas can be computed analytically. Appendix B is devoted to considering  special case of decomposition of $\ad^{\otimes 3}$ into irreducible subrepresentations for the $\sla_n$ Lie algebras by means of Young diagrams, and the correspondence with the universal description is stated. Finally, in Appendix C, we extend the results of \cite{MSV} and find  universal generating functions for the eigenvalues of higher Casimir operators for the representations $X_2,\ Y_2,\ Y_2'$ and $Y_2''$ participating in the universal decomposition $\ad^{\otimes 2}=X_0\oplus X_1\oplus X_2\oplus Y_2\oplus Y_2'\oplus Y_2''$.
\section{Basic definitions\label{basic}}

\subsection{Highest split Casimir operators}

Let $\mathfrak{g}$ be a simple complex
Lie algebra with  the basis
elements $X_a$ and defining relations
\be\label{lialg}
[X_a, \; X_b] = X_d \; X^d_{ab}  \; ,
\ee
 where $X^d_{ab}\equiv (X_a)^d_{\;\; b}$ are the structure constants that
 define the adjoint representation: $\ad^d_b(X_a) = (X_a)^d_{\;\; b}$. The Cartan-Killing  metric ${\sf g}_{ab}$ is defined in the standard way:
\be\label{li04}
   {\sf g}_{ab} \equiv X^{d}_{ac} \, X^{c}_{bd} =
   \Tr(\ad(X_a)\cdot \ad(X_b)) \; .
\ee
Recall that the
 structure constants
 \be
 \label{strX}
 X_{abc} \equiv X^{d}_{ab} \, {\sf g}_{dc} \; ,
 \ee
 are antisymmetric under permutation of indices $(a,b,c)$. For
 simple Lie algebras the Cartan-Killing  metric (\ref{li04}) is invertible
 ${\sf g}^{ab} {\sf g}_{bc} = \delta^a_c$,
 and by using the inverse metric ${\sf g}^{ab}$,
 one can define the quadratic Casimir operator
  \be
 \lb{kaz-c2}
 C = {\sf g}^{ab} \; X_a \cdot X_b \;\; \in \;\;
 {\cal U}(\mathfrak{g}) \; ,
  \ee
  which is the central element in the enveloping algebra ${\cal U}(\mathfrak{g})$. Let $V_\lambda$ be the space
  of the irreducible representation $T_\lambda$
  of $\mathfrak{g}$
  with the highest weight $\lambda$. Thus, we have
  \be
  \lb{eigval}
  C  \, V_\lambda = c_2^{(\lambda)} \, V_\lambda \; ,
  \ee
  where $c_2^{(\lambda)}$ is the value of the Casimir operator
  in the representation $T_\lambda$. We can choose the metric
  in the root space such that (see e.g. \cite{Book2})
   \be
 \lb{kvkaz}
 c_2^{(\lambda)} = (\lambda,\lambda + 2 \, \delta) \; ,
 \ee
 where $\delta$ is the Weil vector of the Lie
 algebra $\mathfrak{g}$.

 Now we introduce the first split Casimir operator
 \be
 \lb{kaz-01}
\hC  = {\sf g}^{ab} X_a \, \otimes \,
  X_b  \;\;\; \in \;\;\;   {\cal U}(\mathfrak{g})^{\otimes \, 2} \; ,
 \ee
 which is related to the operator (\ref{kaz-c2})
 by means of the formula
 \be
 \lb{adCC1}
 \Delta(C) = C \otimes I + I \otimes C + 2 \, \hC  \;\;\;\;\; \Rightarrow
 \;\;\;\;\; \hC = \frac{1}{2} (\Delta(C) - C \otimes I - I \otimes C)   \; ,
 \ee
 where $I \in {\cal U}(\mathfrak{g})$ is the unit operator
 and $\Delta$ is the standard comultiplication
 in ${\cal U}(\mathfrak{g})$:
 \be
 \lb{mrep2}
 \Delta(X_a) = (X_a \otimes I + I \otimes X_a) \; , \;\;\;\;
 \Delta(I) = I \otimes I  \; .
 \ee
 Below we use the highest split
 ($n$-split) Casimir operators
 \be
 \lb{kaz-n}
\hC_{(n)}  := \sum_{i<j}^{n} \hC_{ij}
 \;\;\; \in \;\;\;   {\cal U}(\mathfrak{g})^{\otimes \, n} \; ,
 \ee
 where
 $$
 \hC_{ij} : = {\sf g}^{ab} \left(I ^{\otimes (i-1)}
 \, \otimes \, X_a \, \otimes \,  I^{\otimes (j-i-1)} \, \otimes \,  X_b
 \otimes \,  I^{\otimes (n-j)} \right) \; .
 $$
 The highest split Casimir operator (\ref{kaz-n}) is also related to the
 quadratic Casimir operator (\ref{kaz-c2}) and this relation
 is obtained by means of the $n$-th power
 $\Delta^n$ of the comultiplication (\ref{mrep2}):
 \be
 \lb{adCC2}
 \begin{array}{c}
 \Delta^n(C) \equiv (\Delta \otimes {\rm id}^{\otimes n-1})\Delta^{n-1}(C)
 = \sum\limits_{k=1}^n \, C_k  + 2 \, \sum\limits_{i<j}^{n} \hC_{ij}
 \;\;\;\;\; \Rightarrow
 \;\;\;\;\; \hC_{(n)} = \frac{1}{2} (\Delta^n(C) - \sum\limits_{k=1}^n \, C_k)
 \; , 
  \end{array}
 \ee
 \be
 \lb{adCC3}
 C_k := I ^{\otimes (k-1)}
 \, \otimes \, C \, \otimes \,  I^{\otimes (n-k)} \; .
 \ee
Relations (\ref{adCC2}) generalize formulas (\ref{adCC1}).
Now we act by the operator (\ref{adCC2}) on the expansion
$$
V_{\lambda_1} \otimes V_{\lambda_2} \otimes \cdots
\otimes V_{\lambda_n} = \sum_\lambda V_{\lambda} \; ,
$$
and as a result we obtain
\be
 \lb{adCC5}
\hC_{(n)} \, V_{\lambda}
= \frac{1}{2} (c^{(\lambda)}_2 -
\sum\limits_{k=1}^n \, c^{(\lambda_k)}_2) \, V_{\lambda}
 \ee
where eigenvalues $c^{(\lambda)}_2$ were
defined in (\ref{eigval}) and (\ref{kvkaz}).

The following statement holds (see, for example, \cite{Okubo,Book1,ToHa}).
  \newtheorem{pro1}{Proposition}[subsection]
\begin{pro1}\label{pro1}
The $n$-split Casimir
operator $\hC_{(n)}\in {\cal U}(\mathfrak{g})^{\otimes \, n}$,
 given in (\ref{kaz-n}) and
(\ref{adCC2}), does not depend on
the choice of the basis $X_a$ in $\mathfrak{g}$
and satisfies the condition (which is called $\ad$-invariance or
$\mathfrak{g}$-invariance):
 \be
 \lb{kaz-02}
[\Delta^n(A), \, \hC_{(n)}  ] \equiv
[ \sum_k A_k , \, \hC_{(n)}  ]= 0 \; , \;\;\;\;
\forall A \in \mathfrak{g} \; ,
 \ee
 where $\Delta^n$ is the $n$-th power
 of the comultiplication (\ref{mrep2}) (see the definition in (\ref{adCC2}))
 and the operators $A_k \in {\cal U}(\mathfrak{g})^{\otimes \, n}$
 are defined according to (\ref{adCC3}).
 In addition, the operators $\hC_{ij}$ obeys the equations
 \be
 \lb{kaz-03}
 [\hC _{ij}, \, \hC _{ik} + \hC _{jk} ] = 0 \;\;\; \Rightarrow \;\;\;
[\hC _{ik}, \, \hC _{jk} ]
= \frac{1}{2} \; [\hC _{ij}, \, \hC _{ik} - \hC _{jk} ] \; .
 \ee
  \end{pro1}
  {\bf Proof.} Condition (\ref{kaz-02}) follows
  from the obvious relation $[A, \; C]=0$ and relation
  (\ref{adCC2}). Equations (\ref{kaz-03}) follow
  from (\ref{kaz-02}) for $n=2$ since
  $(\Delta \otimes id ) \hC = \hC_{13} + \hC_{23}$.
  \hfill \qed

Assume an operator $A$ to act on the vector space $V$. If $A$ satisfies an equation of the form
\begin{equation}
(A-a_1)(A-a_2)\cdots (A-a_p)=0
\end{equation}
with all $a_1,\dots,a_p$ different, then $A$ is diagonalizable and the projector $\proj_{a_i}$ onto the eigenspace $V_{a_i}$ of $A$ with the eigenvalue $a_i$ is given by \cite{Book2}
\begin{equation}\label{GenPr}
\proj_{a_i}=\prod_{\substack{j=1\\j\neq i}}^p\frac{A-a_jI}{a_i-a_j},
\end{equation}
where $I$ is the identity operator on $V$. It is clear that the dimension of $V_{a_i}$ coincides with the trace of $\proj_{a_i}$:
\begin{equation}
\dim V_{a_i}=\mathrm{Tr}\proj_{a_i}.
\end{equation}
In the present paper, the role of $A$ will be played by different symmetrizations of the $3$-split Casimir operator taken in the adjoint representation.

  \subsection{Vogel parametrization and $n$-split Casimir
  operators in $\ad$-representation}

  Further we consider the $n$-split Casimir operators in the adjoint
  representation $\ad^{\otimes n}(\hC_{(n)})$ which act in the
  space $V_{\ad}^{\otimes n}$, and to simplify the notation, we
  write $\hC_{(n)}$ instead of $\ad^{\otimes n}(\hC_{(n)})$.
  Introduce the operators
  \be
  \lb{adK1}
   1^{a_1 a_2}_{\;\; b_1 b_2}
  = \delta^{a_1}_{b_1} \delta^{a_2}_{b_2} \; , \;\;\;\;\;
  P^{a_1 a_2}_{\;\; b_1 b_2}
  = \delta^{a_1}_{b_2} \delta^{a_2}_{b_1} \; , \;\;\;\;\;
  K^{a_1 a_2}_{\;\; b_1 b_2}
  = \delta^{a_1 a_2} \delta_{b_1 b_2} \; , \;\;\;\;\;
 \hC_{\pm} = \frac{1}{2} (1 \pm P)\hC  \; \equiv \;
 P_{\pm} \hC  \; .
 \ee
It can be shown that for $\hC_+$, the following identity holds:
\begin{equation}
(\hC_+ + 1) (\hC_+ + \frac{\alpha}{2t})  (\hC_+ + \frac{\beta}{2t})(\hC_+ + \frac{\gamma}{2t})  P_{+} = 0,
\end{equation}
where $t:=\alpha+\beta+\gamma$ is the normalization parameter, and the values of $\alpha$, $\beta$ and $\gamma$ for simple Lie algebras are given in Table 1 (see, e.g., \cite{MSV}). These parameters, which are defined up to permutation and simultaneous multiplication by any real number, were introduced by Vogel in \cite{Fog} and are called the Vogel parameters.
 \begin{center}
	Table 1. \\ [0.2cm]
	\begin{tabular}{|c|c|c|c|c|c|c|c|c|}
		\hline
		$\;\;$ & $\sla_n$ & $\soa_n$ & $\spa_{2r}$ & $\mathfrak{g}_2$ & $\mathfrak{f}_4$
		& $\mathfrak{e}_6$ &
		$\mathfrak{e}_7$ & $\mathfrak{e}_8$  \\
		\hline
		$\frac{\alpha}{2 t}$  &\footnotesize  $-1/n$ &\footnotesize  $-1/(n-2)$ & \footnotesize $1/2(r+1)$ &
		\footnotesize $-1/4$ &\footnotesize  $-1/9$ &\footnotesize  $-1/12$
		&\footnotesize  $-1/18$ &\footnotesize  $-1/30$  \\
		\hline
		$\frac{\beta}{2 t}$  &\footnotesize $1/n$ &\footnotesize $2/(n-2)$ & \footnotesize  $-1/(r+1)$&
		\footnotesize $5/12$ &\footnotesize $5/18$ &\footnotesize $1/4$ &
		\footnotesize $2/9$ &\footnotesize $1/5$ \\
		\hline
		$\frac{\gamma}{2 t}$  &\footnotesize $1/2$ &\footnotesize $(n-4)/2(n-2)$ & \footnotesize  $(r+2)/2(r+1)$ &
		\footnotesize $1/3$ &\footnotesize $1/3$ &\footnotesize $1/3$ &
		\footnotesize $1/3$ &\footnotesize $1/3$ \\
		\hline
	\end{tabular}
\end{center}

    \newtheorem{pro2}[pro1]{Proposition} 
\begin{pro2}\label{pro2}
(see \cite{Fog,Delig,Cvit,IsPr,IsKri}) For the $2$-split Casimir operator $\hC _{(2)} \equiv \hC$ in the $\ad$-representation we obtain the following universal formulas:
 \be
 \lb{idCC}
 \hC_{-}^2 = - \frac{1}{2}  \hC_{-}  \; ,  \;\;\;
 \hC_{-}^{k+1} = \Bigl(- \frac{1}{2}\Bigr)^k  \hC_{-}  \; ,
 \ee
  \be
 \lb{idK}
 \hC_{-} \, K = 0 = K \, \hC_{-}  \; , \;\;\;
 \hC \, K = K \, \hC = - K \; ,
\;\;\;
 \hC_{+} \, K = K \, \hC_{+} = - K \; , \;\;\;
 K^2 = \dimg \, K \; ,
 \ee
  \be
 \lb{iskri01}
 \begin{array}{c}
  {\rm Tr}_i (\hC_{ik}) = 0  , \;\;
  {\rm Tr}_i (\hC_{ik} \hC_{ij}) = - \hC_{jk}  , \;\;
  {\rm Tr}_i (P_{ik} \hC_{ik})
 = (C)_k = I_{(k)}  \;\;  \Rightarrow \;\;
 {\rm Tr}(P_{12}\hC_{12}) = \dimg \;  , \\ [0.2cm]
  {\rm Tr}(\hC_{12}^2) = \dimg , \;\;
   {\rm Tr}(P_{12}(\hC_{12})^2)
   = \frac{1}{2} \dimg , \;\;\;
 {\rm Tr}(\hC_{\pm}) = \pm \frac{1}{2}
 \dimg \, ,
  \end{array}
 \ee
 \be
 \lb{trac1}
    {\rm Tr}(\hC_{-}^{k}) = (- \frac{1}{2})^k
   \dimg , \;\;
   {\rm Tr}(\hC_{+}^2) = {\rm Tr}(\hC_{12}^2 - \hC_{-}^2) =
   \frac{3}{4}  \dimg ,
 \ee
  \be
\lb{chcp4}
\hC_+^3 +\frac{1}{2} \hC_+^2 = \mu_1 \hC_+
 + \mu_2 (1 + P -2 K) \; , \;\;\;
(\hC_+ + 1) (\hC_+ + \frac{\alpha}{2t})
 (\hC_+ + \frac{\beta}{2t})(\hC_+ + \frac{\gamma}{2t})
 P_{+} = 0 \; ,
\ee
 $$
 \hC_+^4 +\frac{1}{2} \hC_+^3 - \mu_1 \hC_+^2
 - 2 \mu_2 \hC_+  = 2 \mu_2 K  \;\; \Rightarrow \;\;
 \hC_+^4 +\frac{3}{2} \hC_+^3 + (\frac{1}{2}- \mu_1) \hC_+^2
 - (2 \mu_2 +\mu_1) \hC_+  = 2 \mu_2 P_{+} .
 $$
where  $C$ is the quadratic Casimir operator (\ref{kaz-c2})
 in the adjoint representation, ${\rm Tr}_i$ is the trace in the $i$-th space,
 ${\rm Tr} :=  {\rm Tr}_{1}{\rm Tr}_{2}$ and
 \be
\lb{abcd04}
\begin{array}{c}
 \displaystyle
\frac{1}{2} \mu_1 + \mu_2 (\dimg -1 )
= \frac{1}{4} \; , \;\;\;\;\;\;
\mu_1 = - \frac{\alpha\beta  + \alpha\gamma  + \beta\gamma}{4t^2}
 \; , \quad
\mu_2 = - \frac{\alpha\beta \gamma}{16 t^3} \; , \;\;\;
t := \alpha + \beta + \gamma
\;\;\;\; \Rightarrow \\ [0.3cm]
\displaystyle
 \dimg = 1 +
 \frac{\bigl(1/2 -\mu_1 \bigr)}{2\mu_2}
 = \frac{(\alpha-2t)(\beta-2t)(\gamma-2t)}{
 \alpha\beta\gamma}\; .
\end{array}
\ee

The dimensions of the invariant subspaces in $P_{-} V_{\ad}^{\otimes 2}$ are
\begin{align}
& \dim V^{(-)}_{(0)}=\Tr P^{(-)}_{(0)}=\dimg\, ,\label{ad2mdim1}\\
& \dim V^{(-)}_{(-\frac{1}{2})}=\Tr P^{(-)}_{(-\frac{1}{2})}=\frac{1}{2}\dimg(\dimg-3)\, \label{ad2mdim2},
\end{align}
where
$P_{-} = P^{(-)}_{(0)}+P^{(-)}_{(-\frac{1}{2})}$
and $P^{(-)}_{(\lambda)}$
are the projectors (constructed
 from the characteristic identity on the left of (\ref{idCC})) on the eigen-spaces of $\hC_{-}$ with
the eigenvalues $\lambda$.

The dimensions of the invariant subspaces in
$P_{+} V_{\ad}^{\otimes 2}$ are
  \bea
  &&\dim V_{(-1)} =
 \Tr P^{(+)}_{(-1)}  = 1 \\
  &&\dim V_{(-\frac{\alpha}{2t})} =
 {\Tr} \, P^{(+)}_{(-\frac{\alpha}{2t})}
 =-\frac{(3\alpha-2t)(\beta-2t)(\gamma-2t)t(\beta+t)(\gamma+t)}{
\alpha^2(\alpha-\beta)\beta(\alpha-\gamma)\gamma} \, , \lb{dim02a}\\
 \lb{dim02b}
  &&\dim V_{(-\frac{\beta}{2t})} =
 {\Tr} \, P^{(+)}_{(-\frac{\beta}{2t})}
 =-\frac{(3\beta-2t)(\alpha-2t)(\gamma-2t)t(\alpha+t)(\gamma+t)}{
\beta^2(\beta-\alpha)\alpha(\beta-\gamma)\gamma} \, ,\\
&& \dim V_{(-\frac{\gamma}{2t})} =
{\Tr} \, P^{(+)}_{(-\frac{\gamma}{2t})}
 = -\frac{(3\gamma-2t)(\beta-2t)(\alpha-2t)t(\beta+t)(\alpha+t)}{
\gamma^2(\gamma-\beta)\beta(\gamma-\alpha)\alpha}  \, . \lb{dim02c}
\eea
where
$P_{+} = P^{(+)}_{(-1)}+P^{(+)}_{(-\frac{\alpha}{2t})}+
P^{(+)}_{(-\frac{\beta}{2t})}+P^{(+)}_{(-\frac{\gamma}{2t})}$
and $P^{(+)}_{(\lambda)}$
are the projectors (constructed
 from the characteristic identities (\ref{chcp4})) on the eigen-spaces of $\hC_{+}$ with
the eigenvalues $\lambda$.
 \end{pro2}
 {\bf Remark.} For certain special algebras  in Table 1 we have zeros and poles
 in expressions (\ref{dim02a}) --  (\ref{dim02c}). These algebras require 
 special analysis (see \cite{IsPr,IsKri})

\subsection{The $3$-split Casimir operator}

We start with the adjoint representation of a compact Lie algebra $\mathfrak{g}$
with the generators $X_a$ normalized as
\be\label{norm1}
 \Tr \left( \ad(X_a) \cdot \ad(X_b) \right) = - \delta_{ab}.
\ee
The matrix $\hC_{b_1 b_2 b_3}^{a_1 a_2 a_3} :=
 (\hC_{(3)})_{b_1 b_2 b_3}^{a_1 a_2 a_3}$
of the cubic split Casimir operator $\hC_{(3)}$ has the
 form (see eq. (\ref{kaz-n}))
\be\label{02}
 \begin{array}{c}
\hC_{b_1 b_2 b_3}^{a_1 a_2 a_3} = -\sum\limits_{c} \Bigl(
\left( X_c\right)^{a_1}{}_{b_1} \; \left( X_c\right)^{a_2}{}_{b_2} \delta^{a_3}{}_{b_3} +
\left( X_c\right)^{a_1}{}_{b_1} \; \left( X_c\right)^{a_3}{}_{b_3} \delta^{a_2}{}_{b_2}+
\left( X_c\right)^{a_2}{}_{b_2} \; \left( X_c\right)^{a_3}{}_{b_3} \delta^{a_1}{}_{b_1} \Bigr) = \\ [0.2cm]
= (\widehat{C}_{12} + \widehat{C}_{13}
+ \widehat{C}_{23})_{b_1 b_2 b_3}^{a_1 a_2 a_3}
 \; ,
\end{array}
\ee
and acts in the space $V_{\ad}^{\otimes 3}$
 of the representation $\ad^{\otimes 3}$. It is clear that
the matrix is symmetric under any permutations of the columns of indices
$\hC_{b_1 b_2 b_3}^{a_1 a_2 a_3} = \hC_{b_2 b_1 b_3}^{a_2 a_1 a_3} =
\hC_{b_1 b_3 b_2}^{a_1 a_3 a_2}$.
We also need the parts $\aA$, $\sS$ and $\rS$
of the split Casimir operator $\hC_{(3)}$
symmetrized according to the Young diagrams $[3]$, $[1^3]$, $[2,1]$ and $[2,1]'$,
and defined as follows:
\be\label{03}
\begin{array}{c}
\hC_{[3]} =	\sS = \mathbb{S}_{[3]} \cdot  \hC_{(3)} = \hC_{(3)} \cdot \mathbb{S}_{[3]}
	\;\; \Rightarrow \;\; \\ [0.2cm]
	\sS_{b_1 b_2 b_3}^{a_1 a_2 a_3}= \frac{1}{6} \left( \hC_{b_1 b_2 b_3}^{a_1 a_2 a_3}+\hC_{b_2 b_3 b_1}^{a_1 a_2 a_3}+\hC_{b_3 b_1 b_2}^{a_1 a_2 a_3}+\hC_{b_1 b_3 b_2}^{a_1 a_2 a_3}+\hC_{b_3 b_2 b_1}^{a_1 a_2 a_3}+\hC_{b_2 b_1 b_3}^{a_1 a_2 a_3}\right),\\ [0.3cm]
\hC_{[1^3]} =	\aA  = \mathbb{S}_{[1^3]} \cdot  \hC_{(3)} = \hC_{(3)} \cdot \mathbb{S}_{[1^3]}
	\;\; \Rightarrow \;\; \\ [0.2cm]
	\aA_{b_1 b_2 b_3}^{a_1 a_2 a_3 }=
	\frac{1}{6} \left( \hC_{b_1 b_2 b_3}^{a_1 a_2 a_3}+
	\hC_{b_2 b_3 b_1}^{a_1 a_2 a_3} +\hC_{b_3 b_1 b_2}^{a_1 a_2 a_3}-
	\hC_{b_1 b_3 b_2}^{a_1 a_2 a_3} -\hC_{b_3 b_2 b_1}^{a_1 a_2 a_3}
	-\hC_{b_2 b_1 b_3}^{a_1 a_2 a_3}\right),
\end{array}
\ee
\be\label{04}
\begin{array}{c}
\hC_{[2,1]'} = \rS =
\mathbb{S}_{[2,1]'} \cdot  \hC_{(3)} = \hC_{(3)} \cdot \mathbb{S}_{[2,1]'}
\;\; \Rightarrow \;\; \\ [0.3cm]
\rS_{b_1 b_2 b_3}^{a_1 a_2 a_3 }=
\frac{1}{3} \left( \, \hC_{b_1 b_2 b_3}^{a_1 a_2 a_3}
+\hC_{b_2 b_1 b_3}^{a_1 a_2 a_3}
-\hC_{b_3 b_2 b_1}^{a_1 a_2 a_3}-\hC_{b_2 b_3 b_1}^{a_1 a_2 a_3}\right),\\[0.3cm]
\hC_{[2,1]''} = \rS' =
\mathbb{S}_{[2,1]''} \cdot  \hC_{(3)} = \hC_{(3)} \cdot \mathbb{S}_{[2,1]''}
\;\; \Rightarrow \;\; \\ [0.3cm]
(\rS')_{b_1 b_2 b_3}^{a_1 a_2 a_3 }=
\frac{1}{3} \left( \, \hC_{b_1 b_2 b_3}^{a_1 a_2 a_3}
+\hC_{b_3 b_2 b_1}^{a_1 a_2 a_3}
-\hC_{b_2 b_1 b_3}^{a_1 a_2 a_3}-\hC_{b_3 b_1 b_2}^{a_1 a_2 a_3}\right),
\end{array}
\ee
where we used primitive Young symmetrizers (see e.g. \cite{Book2})
\be
\begin{gathered}
\label{04ii}
\mathbb{S}_{[3]} =
\frac{1}{3!} (I + P_{23} + P_{12}P_{23})(I + P_{12}) \; , \;\;\;\;
\mathbb{S}_{[1^3]}
= \frac{1}{3!} (I - P_{23} + P_{12}P_{23})(I - P_{12}) \; ,\\
\mathbb{S}_{[2,1]'} =  \frac{1}{3}
(I+P_{12})(I-P_{13}) \; , \;\;\;\; \mathbb{S}_{[2,1]''} =  \frac{1}{3}
(I+P_{13})(I-P_{12}) \; ,
\end{gathered}
\ee
and $I$ is the unit matrix in $V_{\ad}^{\otimes 3}$. Since $\mathbb{S}_{[2,1]''}=P_{23}\,\mathbb{S}_{[2,1]'}P_{23}$ and $\hC_{(3)}$ commutes with $P_{23}$, any relation for $\hC_{[2,1]''}$ can be readily obtained from an analogous relation for $\hC_{[2,1]'}$ by substituting $\hC_{[2,1]'}$ with $\hC_{[2,1]''}$. Therefore, we will not specifically consider the case of the $[2,1]''$ part of the $3$-split Casimir operator $\hC_{(3)}$. The central idempotents for
th permutation group algebra $\mathbb{C}[S_3]$ are
$$
\mathbb{S}_{[3]}  \; , \;\;\;\;\;\;
\mathbb{S}_{[1^3]} \; , \;\;\;\;\;\;
\mathbb{S}_{[2,1]} =  \mathbb{S}_{[2,1]'}
 + \mathbb{S}_{[2,1]''} =  1 - \mathbb{S}_{[3]}
 - \mathbb{S}_{[1^3]}  \; .
$$

\subsection{The $\kK$ operator
in $V_{\ad}^{\otimes 3}$}

Further we also need the following operator:
\be\label{05i}
\begin{array}{c}
\kK_{b_1 b_2 b_3}^{a_1 a_2 a_3} = \delta_{b_1 b_2} \delta^{a_1 a_2} \delta_{b_3}^{a_3}+\delta_{b_1 b_3} \delta^{a_1 a_3} \delta_{b_2}^{a_2}+\delta_{b_2 b_3} \delta^{a_2 a_3} \delta_{b_1}^{a_1} =
 (K_{12} + K_{13} + K_{23})_{b_1 b_2 b_3}^{a_1 a_2 a_3}  ,
\end{array}
\ee
together with its symmetrized parts
\begin{gather}\label{06}
(\mathbb{S}_{[3]} \cdot \mathbb{K})_{b_1 b_2 b_3}^{a_1 a_2 a_3 } =
(\kK_{[3]})_{b_1 b_2 b_3}^{a_1 a_2 a_3 }=
\frac{1}{6} \left( \kK_{b_1 b_2 b_3}^{a_1 a_2 a_3}+\kK_{b_2 b_3 b_1}^{a_1 a_2 a_3}+\kK_{b_3 b_1 b_2}^{a_1 a_2 a_3}+\kK_{b_1 b_3 b_2}^{a_1 a_2 a_3}+\kK_{b_3 b_2 b_1}^{a_1 a_2 a_3}+\kK_{b_2 b_1 b_3}^{a_1 a_2 a_3}\right) = \nn \\[-0.1cm] 
= \frac{1}{3} \left(\kK + P_{12} (K_{13} + K_{23})
+ P_{23} (K_{12} + K_{13}) + P_{13} (K_{12} + K_{23})
\right)_{b_1 b_2 b_3}^{a_1 a_2 a_3} , \;\;\;
 \\[0.1cm] 
\kK_{[2,1]} := \mathbb{S}_{[2,1]} \cdot \mathbb{K} = \kK-\kK_{[3]} \nn.
\end{gather}
It is clear that $\mathbb{S}_{[1^3]}  \cdot \mathbb{K} = 0
= \mathbb{K} \cdot \mathbb{S}_{[1^3]}$ and
\be
\label{perm}
\begin{array}{c}
P_k \cdot  \kK = \kK \cdot  P_k \;\;\;\;\;\;
 \Rightarrow \;\;\;\;\;\;
\mathbb{S}_{[3]}  \cdot \mathbb{K} = \mathbb{K} \cdot \mathbb{S}_{[3]} \; ,
\;\;\; \mathbb{S}_{[2,1]}  \cdot \mathbb{K} = \mathbb{K} \cdot \mathbb{S}_{[2,1]} \; ,
\end{array}
\ee
where $P_k = P_{k k+1}$ $(k=1,2)$.
Then we deduce
$\sI_{[3]} \cdot \kK_{[3]} = \kK_{[3]} \cdot \sI_{[3]} = \kK_{[3]}$ and therefore we have
\be
\label{sk2}
\begin{array}{c}
\kK_{[3]}^2 = (\mathbb{S}_{[3]} \cdot \mathbb{K})^2 =
\mathbb{S}_{[3]} \cdot (K_{12} + K_{13} + K_{23})^2 = \\ [0.3cm]
 = \mathbb{S}_{[3]} \cdot (K_{12}^2 + K_{13}^2 + K_{23}^2 +
[K_{12},K_{13}]_+ + [K_{12},K_{23}]_+ + [K_{13},K_{23}]_+) = \\ [0.3cm]
 = \mathbb{S}_{[3]} \cdot \Bigl(\dimg (K_{12} + K_{13} + K_{23}) +
 2 (K_{12} + K_{13} + K_{23}) \Bigr) = (\dimg +2)\; \kK_{[3]} \; ,
\end{array}
 \ee
where we applied the identities  $K_{i j}^2= \dimg K_{i j}$,
$\;K_{i j} K_{k j} = P_{i k} K_{k j}$ and
$\mathbb{S}_{[3]} \cdot P_{i k}  = \mathbb{S}_{[3]}$.
In addition, we obtain
\be
\label{rk2}
\kK_{[2,1]}^2 = (\kK - \kK_{[3]})^2 = \kK^2  - \kK_{[3]}^2 =
(\dimg -1) \, \kK - (\dimg -1) \, \kK_{[3]}
=(\dimg -1) \, \kK_{[2,1]} \; ,
\ee
where we take into account (\ref{sk2}) and the identity
$
\kK^2 = (\dimg -1) \, \kK + 3 \, \kK_{[3]}$,
which follows from (\ref{06}). From (\ref{sk2}) and
(\ref{rk2}) we see that the operators
\be
\lb{K3K12}
\frac{1}{(\dimg+2)} \kK_{[3]}
\, , \;\;\;\;\;\;
\frac{1}{(\dimg-1)} \kK_{[2,1]}
\ee
are the invariant projectors on the spaces
$\mathbb{S}_{[3]} V_{\ad}^{\otimes 3}$ and $\mathbb{S}_{[2,1]} V_{\ad}^{\otimes 3}$, respectively.

\noindent
{\bf Remark 1.}
By using relations (\ref{06}) and the identities
 ${\rm Tr}_{_{123}}(\kK) = 3 \dimg^2$ and ${\rm Tr}_{_{123}}(P_{ij} K_{kj})= \\dimg$ $(i\neq k)$, we find
$$
{\rm Tr}_{_{123}}(\kK_{[3]}) = \dimg^2 + 2 \dimg
\; , \;\;\;
{\rm Tr}_{_{123}}(\kK_{[2,1]}) =
2 \bigl(\dimg^2 -\dimg\bigr)  \; ,
$$
and, in view of (\ref{sk2}) and (\ref{rk2}), we have the traces of the
projectors (\ref{K3K12})
\be
\label{dsrk}
\frac{1}{(\dimg+2)} {\rm Tr}_{_{123}}(\kK_{[3]}) = \dimg
\, , \;\;\;\;\;\;
\frac{1}{(\dimg-1)} {\rm Tr}_{_{123}}(\kK_{[2,1]}) =
 2 \dimg \; ,
\ee
which give the dimensions of the corresponding
subspaces in $\mathbb{S}_{[3]} (V_{\ad}^{\otimes 3})$
and $\mathbb{S}_{[2,1]} (V_{\ad}^{\otimes 3})$.

\noindent
{\bf Remark 2.}
In the expansion of the antisymmetrizer
$\Pi_3 \equiv \mathbb{S}_{[1^3]}$ we always have the rank-1 projector $\mathbb{X}$
\be
\label{dsrk2}
\mathbb{X}_{b_1 b_2 b_3}^{a_1 a_2 a_3} :=
\frac{-1}{\dimg} \,
X^{a_1 a_2 a_3} \; X_{b_1 b_2 b_3} \; ,
\ee
where $X^{a_1 a_2 a_3}$ and $X_{b_1 b_2 b_3}$ are the structure
constants of $\mathfrak{g}$ (see e.g. (\ref{strX}))
and we add here the normalization factor $\frac{-1}{\dimg}$
 in view of the identity
 $$
 X^{a_1 a_2 a_3} \; X_{a_1 a_2 a_3} = - \dimg
 \;\; \Rightarrow \;\;
 \Tr_{123} \mathbb{X} = 1  \; .
 $$
 Thus, indeed, the projector $\mathbb{X}$ extracts 
 the one dimensional subspace (singlet) from
 the space $\mathbb{S}_{[1^3]} (V_{\ad}^{\otimes 3})$.
As we show below, the operator $\mathbb{X}$ coincides
with the projector $P^{(-)}_{(-\frac{3}{2})}$ on the eigenspace of
$\hC_{[1^3]}$ in
$\mathbb{S}_{[1^3]} (V_{\ad}^{\otimes 3})$ with the eigenvalue $(-3/2)$
(see Section {\bf \ref{antsym}}).

\setcounter{equation}0
\section{Universal characteristic identity for the anti-symmetric part of the
$3$-split Casimir operator $\hC_{(3)}$\label{antsym}}

One may check that the anti-symmetric part of the third power of the split Casimir operator obeys the following characteristic identity:
\be
\aA \left( \aA+\frac{1}{2} \right) \left(\aA+\frac{3}{2}\right)\left(\aA+\frac{1}{2}+\hat{\alpha}\right) \left( \aA +\frac{1}{2}+\hat{\beta}\right)\left( \aA +\frac{1}{2}+\hat{\gamma}\right)\mathbb{S}_{[1^3]} =0
\ee
where
$$ \hat \alpha = \frac{\alpha}{2t}, \; \hat \beta = \frac{\beta}{2t}, \;\hat \gamma = \frac{\gamma}{2t} .
$$

In the cubic case, the traces of $\aA$ have the generic expressions
(to obtain these traces, we partially use formulas
from Appendix {\bf A})
\bea
\Tr(\mathbb{S}_{[1^3]}) & = & \frac{1}{6} \dimg \left(\dimg-1\right) \left(\dimg-2\right),\nn \\
\Tr(\aA) & = & -\frac{1}{2} \dimg \left(\dimg-2\right),\nn \\
\Tr(\aA^2) & = & \frac{1}{4} \dimg^2\nn \\
\Tr(\aA^3) & = & -\frac{1}{8} \dimg \left( \dimg +6\right),\label{TrAC} \\
\Tr(\aA^4) & = & \frac{1}{16}  \dimg \left(\dimg +24 +12   \left(\dimg -3\right)\ha \hb \hg\right) , \nn \\
\Tr(\aA^5) & = & -\frac{1}{32}  \dimg \left(\dimg+78+76 \left(\dimg-3\right) \ha \hb \hg \right) \nn .
\eea

The projectors onto the irreps corresponding to the eigenvalues of $\aA$ are obtained by \eqref{GenPr} where
\begin{equation}
\begin{gathered}
A=\aA,\ \ I=\mathbb{S}_{[1^3]},\ \ p=6,\\
 a_1=0,\ \ a_2=-\frac{1}{2},\ \ a_3=-\frac{3}{2},\ \ a_4=-\frac{1}{2}-\ha,\ \ a_5=-\frac{1}{2}-\hb,\ \ a_6=-\frac{1}{2}-\hg,
\end{gathered}
\end{equation}
which results in
\begin{equation}\label{aprojs}
\begin{aligned}
P^{[1^3]}_{(0)}&=\frac{4\left(\aA+\frac{1}{2}\right)\left(\aA+\frac{3}{2}\right)\left(\aA+\frac{1}{2}+\ha\right)\left(\aA+\frac{1}{2}+\hb\right)\left(\aA+\frac{1}{2}+\hg\right)\mathbb{S}_{[1^3]}}{3\left(\frac{1}{2}+\ha\right)\left(\frac{1}{2}+\hb\right)\left(\frac{1}{2}+\hg\right)},\\
P^{[1^3]}_{(-\frac{1}{2})}&=-\frac{2\aA\left(\aA+\frac{3}{2}\right)\left(\aA+\frac{1}{2}+\ha\right)\left(\aA+\frac{1}{2}+\hb\right)\left(\aA+\frac{1}{2}+\hg\right)}{\ha\hb\hg},\\
P^{[1^3]}_{(-\frac{3}{2})}&=\frac{2\aA\left(\aA+\frac{1}{2}\right)\left(\aA+\frac{1}{2}+\ha\right)\left(\aA+\frac{1}{2}+\hb\right)\left(\aA+\frac{1}{2}+\hg\right)}{3(\ha-1)(\hb-1)(\hg-1)},\\
P^{[1^3]}_{(-\frac{1}{2}-\ha)}&=-\frac{\aA\left(\aA+\frac{1}{2}\right)\left(\aA+\frac{3}{2}\right)\left(\aA+\frac{1}{2}+\hb\right)\left(\aA+\frac{1}{2}+\hg\right)}{\ha\left(\ha+\frac{1}{2}\right)\left(\ha-1\right)(\ha-\hb)(\ha-\hg)},\\
P^{[1^3]}_{(-\frac{1}{2}-\hb)}&=-\frac{\aA\left(\aA+\frac{1}{2}\right)\left(\aA+\frac{3}{2}\right)\left(\aA+\frac{1}{2}+\hg\right)\left(\aA+\frac{1}{2}+\ha\right)}{\hb\left(\hb+\frac{1}{2}\right)\left(\hb-1\right)(\hb-\hg)(\hb-\ha)},\\
P^{[1^3]}_{(-\frac{1}{2}-\hg)}&=-\frac{\aA\left(\aA+\frac{1}{2}\right)\left(\aA+\frac{3}{2}\right)\left(\aA+\frac{1}{2}+\ha\right)\left(\aA+\frac{1}{2}+\hb\right)}{\hg\left(\hg+\frac{1}{2}\right)\left(\hg-1\right)(\hg-\ha)(\hg-\hb)}.
\end{aligned}
\end{equation}
The dimensions of the respective subrepresentations equal the traces of \eqref{aprojs} and can be computed using \eqref{TrAC}. The results read
\bea
\dim_0 & =  X_3+X_3' =& \frac{1}{6}\dimg \left( \dimg-1\right)\left( \dimg-8\right), \nn \\
\dim_{-\frac{1}{2}} & = X_2  = & \frac{1}{2}\dimg \left( \dimg -3\right), \nn\\
\dim_{-\frac{3}{2}} & = X_0 = & 1, \nn \\
\dim_{-\hat{\alpha}-\frac{1}{2}} & =  Y_2 =& -\frac{(3\ha-1)(\hb-1)(\hg-1)(2\hb+1)(2\hg+1)}{
8 \ha^2 (\ha-\hb)(\ha-\hg)\hb\hg}, \label{uni01} \\
\dim_{-\hat{\beta}-\frac{1}{2}} & =  Y_2' =& -\frac{(3\hb-1)(\hg-1)(\ha-1)(2\hg+1)(2\ha+1)}{
	8 \hb^2 (\hb-\hg)(\hb-\ha)\hg\ha}, \nn \\
\dim_{-\hat{\gamma}-\frac{1}{2}} & =  Y_2'' =& -\frac{(3\hg-1)(\ha-1)(\hb-1)(2\ha+1)(2\hb+1)}{
	8 \hg^2 (\hg-\ha)(\hg-\hb)\ha\hb}.\nn
\eea
The notation $X_0,\ X_1,\ X_2,\ X_3,\ X_3',\ Y_2,\ Y_2',\ Y_2''$ etc. for subrepresentations entering the universal decompositions of $\ad^{\otimes k}$ was introduced in \cite{Fog} and \cite{Delig}.
The full dimension of all the representations is
\be
\frac{1}{6} \dimg \left(\dimg-1\right)\left(\dimg-2\right),
\ee
as it should be.

Thus,
\be
\wedge^3 {\cal L}=X_0\oplus X_2 \oplus \left(X_3 \oplus X_3'\right)\oplus  Y_2 \oplus Y_2' \oplus Y_2'' .
\ee

\subsection{The case with reduced characteristic identity }
For  algebras of the classical series, a full characteristic identity is needed.
\begin{itemize}
	\item for the exceptional algebras $Y_2''=0$ and the characteristic identity acquires the form
		$$ \aA \left( \aA+\frac{1}{2} \right) \left(\aA+\frac{3}{2}\right)\left(\aA+\frac{1+2\,\ha}{2}\right) \left( \aA +\frac{2-3\,\ha}{3}\right)\mathbb{S}_{[1^3]} =0 .$$ 
\end{itemize}

\setcounter{equation}0
\section{Universal characteristic identity for the symmetric part of the  $3$-split Casimir operator}

One may check that the symmetric part of the third power of the split Casimir operator obey the following characteristic identity:
\be\label{sS}
 \left( \sS+\frac{1}{2} \right) \left(\sS+1\right)\left(\sS+\frac{1}{2}-\hat{\alpha}\right) \left( \sS +\frac{1}{2} - \hat{\beta}\right)\left( \sS +\frac{1}{2} - \hat{\gamma}\right) \left(\sS+3 \hat{\alpha}\right)\left(\sS+3 \hat{\beta}\right)\left(\sS+3 \hat{\gamma}\right)\mathbb{S}_{[3]} =0 .
\ee

In the cubic case, the traces of $\sS$ have the generic expressions (to obtain these traces, we partially use formulas
from Appendix {\bf A})
\bea
\Tr(\mathbb{S}_{[3]}) & = & \frac{1}{6} \dimg \left(\dimg+1\right) \left(\dimg+2\right),\nn \\
\Tr(\sS) & = & \frac{1}{2} \dimg \left(\dimg+2\right),\nn \\
\Tr(\sS^2) & = & \frac{1}{4} \dimg \left( 3\, \dimg+16\right),\nn \\
\Tr(\sS^3) & = & -\frac{1}{8} \dimg \left( \dimg-58\right),\nn \\
\Tr(\sS^4) & = & \frac{1}{16}  \dimg \left(7\,\dimg +184 +12   \left(\dimg -3\right)\left(\dimg+17\right)\ha \hb \hg\right) , \label{TrSC}\\
\Tr(\sS^5) & = & -\frac{1}{32}  \dimg \left(9\,\dimg-562+4 \left(\dimg-3\right)\left(4\, \dimg-301\right) \ha \hb \hg \right), \nn\\
\Tr(\sS^6) & = &\frac{1}{64}  \dimg \left(23\,\dimg+1696+ 4 \left(\dimg-3\right)\left( 17\, \dimg+1630\right) \ha \hb \hg+\right. , \nn \\
&&\left. 48 \left(\dimg-3\right)  \left(\dimg-1\right) \left(\dimg+137\right) \ha^2 \hb^2 \hg^2 \right) .\nn \\
\Tr(\sS^7) & = &-\frac{1}{128} \dimg \left(41\,\dimg-5098+ 8 \left(\dimg-3\right)\left( 17\, \dimg -2637\right) \ha \hb \hg+\right. \nn \\
&&\left. 16 \left(\dimg-3\right)  \left(7\, \dimg^2-1325\, \dimg+4450\right) \ha^2 \hb^2 \hg^2 \right)\nn
\eea

The dimensions of the irreps corresponding to the eigenvalues of the $\sS$ are obtained by calculating the respective projectors by \eqref{GenPr}, where
\begin{equation}
\begin{gathered}
A=\sS,\ \ I=\mathbb{S}_{[3]},\ \ p=8,\\
 a_1=-\frac{1}{2},\ a_2=-1,\ a_3=-\frac{1}{2}+\ha,\ a_4=-\frac{1}{2}+\hb,\ a_5=-\frac{1}{2}+\hg,\ a_6=-3\ha,\ a_7=-3\hb,\ a_8=-3\hg
\end{gathered}
\end{equation}
and using \eqref{TrSC}. The results read
\bea\label{uniDimsSym}
\dim_{-\frac{1}{2}} & = X_2  = & \frac{1}{2}\,\dimg \left( \dimg -3\right), \nn\\
\dim_{-1} & = 2 X_1 = & 2\, \dimg, \nn \\
\dim_{\hat{\alpha}-\frac{1}{2}} & =  B =& \frac{ (\ha-1)(\hb-1)(\hg-1)(2\,\ha+\hb)(2\,\ha+\hg)(2\, \hb+1)(2\, \hg+1)(3\, \hb-1)(3\,\hg-1)}{8\, \ha^2\,(\ha-\hb)(\ha-\hg)(2\, \hb-\hg)(2\hg-\hb)\,\hb^2\,\hg^2},\nn \\
\dim_{\hat{\beta}-\frac{1}{2}} & =  B' =& \frac{ (\hb-1)(\hg-1)(\ha-1)(2\,\hb+\hg)(2\,\hb+\ha)(2\, \hg+1)(2\, \ha+1)(3\, \hg-1)(3\,\ha-1)}{8\, \hb^2\,(\hb-\hg)(\hb-\ha)(2\, \hg-\ha)(2\ha-\hg)\,\hg^2\,\ha^2},\nn  \\
\dim_{\hat{\gamma}-\frac{1}{2}} & = B'' =& \frac{ (\hg-1)(\ha-1)(\hb-1)(2\,\hg+\ha)(2\,\hg+\hb)(2\, \ha+1)(2\, \hb+1)(3\, \ha-1)(3\,\hb-1)}{8\, \hg^2\,(\hg-\ha)(\hg-\hb)(2\, \ha-\hb)(2\hb-\ha)\,\ha^2\,\hb^2},\nn \\
\dim_{- 3 \ha} & = Y_3 = &  -\frac{(\ha-1)(5\, \ha-1)(\hb-1)(\hg-1)(2\hb+1)(2\, \hg+1)(2\,\hb +\hg)(2\, \hg+\hb)}{24\, \ha^3\,(\ha-\hb)(\ha-\hg)(2\, \ha-\hb)(2\,\ha -\hg)\,\hb\,\hg}, 
\label{uni02} \\
\dim_{- 3 \hb} & = Y_3' = &  -\frac{(\hb-1)(5\, \hb-1)(\hg-1)(\ha-1)(2\hg+1)(2\, \ha+1)(2\,\hg +\ha)(2\, \ha+\hg)}{24\, \hb^3\,(\hb-\hg)(\hb-\ha)(2\, \hb-\hg)(2\,\hb -\ha)\,\hg\,\ha}, \nn \\
\dim_{- 3 \hg} & = Y_3'' = &  -\frac{(\hg-1)(5\, \hg-1)(\ha-1)(\hb-1)(2\ha+1)(2\, \hb+1)(2\,\ha +\hb)(2\, \hb+\ha)}{24\, \hg^3\,(\hg-\ha)(\hg-\hb)(2\, \hg-\ha)(2\,\hg -\hb)\,\ha\,\hb}\nn.
\eea
The full dimension of  all representations is
\be
\frac{1}{6} \dimg \left(\dimg+1\right)\left(\dimg+2\right),
\ee
as it should be.

Thus,
\be
{\cal S}^3 {\cal L}=2 X_1\oplus X_2 \oplus B \oplus B'\oplus B'' \oplus  Y_3 \oplus Y_3' \oplus Y_3'' .
\ee

\subsection{The cases with reduced characteristic identity }
The full characteristic identity \p{sS} is needed for the $\sla_n$ algebras. In the remaining cases, it reduces as follows:
\begin{itemize}
	\item for the algebras $\soa_n$ and $\spa_{2r}$ we have $B=0$ and $Y_3''=0$ and, therefore, the characteristic identities for $\soa_n$ and $\spa_{2r}$ read, respectively,
$$
 \left( \sS+\frac{1}{2} \right) \left(\sS+1\right)\left( \sS -\frac{3}{n-2} \right)\left( \sS +\frac{1}{n-2} \right) \left(\sS+ \frac{6}{n-2} \right)\left(\sS+\frac{n-6}{2(n-2)}\right)\mathbb{S}_{[3]} =0 ,
$$
and
$$
\left( \sS+\frac{1}{2} \right) \left(\sS+1\right)\left( \sS +\frac{r+3}{2(r+1)} \right)\left( \sS -\frac{1}{2(r+1)} \right) \left(\sS+ \frac{3}{2(r+1)} \right)\left(\sS-\frac{3}{r+1}\right)\mathbb{S}_{[3]} =0.
$$

	\item for the exceptional algebras $B=B'=0$ and $(\sS+3 \hg)=(\sS+1)$,
	so the multiplier $(\sS+1)$ appears twice in the identity. Moreover, one may check that $\dim\,(Y_3')=-\dimg<0$. This is another indication of that using the split Casimir operator we cannot distinguish between the representations $X_1$ and $Y_3''$ in the case of exceptional Lie algebras. Thus, omitting one multiplier $(\sS+1)$, we can simplify the characteristic identity to
$$ \left( \sS+\frac{1}{2} \right) \left(\sS+1\right)\left(\sS+\frac{1}{6}\right)\left(\sS+\frac{1- 6\, \ha}{2}\right)\left(\sS+3 \ha\right)\mathbb{S}_{[3]} =0 .$$
Note that in this identity $\dim_{-1} = \dimg$.
\end{itemize}

\setcounter{equation}0
\section{Universal characteristic identity for the hook parts of the $3$-split Casimir operator}

One may check that the hook part of the third power of the split Casimir operator obeys the following characteristic identity:
\be\label{rS}
\begin{aligned}
&\left( \rS+\frac{1}{2} \right) \left(\rS+1\right)\left(\rS+\frac{1}{2}-\hat{\alpha}\right) \left( \rS +\frac{1}{2} - \hat{\beta}\right)\left( \rS +\frac{1}{2} - \hat{\gamma}\right) \times \\ &\left(\rS+\frac{1}{2}+\hat{\alpha}\right) \left( \rS +\frac{1}{2} + \hat{\beta}\right)\left( \rS +\frac{1}{2} + \hat{\gamma}\right) \left(\rS+\frac{3}{2} \hat{\alpha}\right)\left(\rS+\frac{3}{2}  \hat{\beta}\right)\left(\rS+\frac{3}{2}  \hat{\gamma}\right)\mathbb{S}_{[2,1]'} =0 .
\end{aligned}
\ee

The traces of $\rS$ have the generic expressions
(to obtain these traces we use formulas
from Appendix {\bf A})
\bea
\Tr(\mathbb{S}_{[2,1]'}) & = & \frac{1}{3} \dimg \left(\dimg-1\right) \left(\dimg+1\right),\nn \\
\Tr(\rS) & = & - \dimg,\nn \\
\Tr(\rS^2) & = & \dimg\left(\dimg-2\right)\nn \\
\Tr(\rS^3) & = & -\frac{1}{4} \dimg \left( \dimg +1\right),\nn \\
\Tr(\rS^4) & = & \frac{1}{4}\dimg \left(2\left(\dimg-1\right)+3\left(\dimg-3\right)\left(\dimg-9\right)\ha\hb\hg\right) , \nn \\
\Tr(\rS^5) & = & -\frac{1}{16} \dimg \left(\left(5\dimg+1\right)+2\left(\dimg-3\right)\left(4\dimg-39\right)\ha\hb\hg\right) \nn \\
\Tr(\rS^6) & = &\frac{1}{16} \dimg \left(2\left(3\dimg-1\right)+\left(\dimg-3\right)\left(17\dimg-228\right)\ha\hb\hg+\right.\nn \\
&&\left. 12\left(\dimg-1\right)\left(\dimg-3\right)\left(\dimg-24\right)\ha^2\hb^2\hg^2\right),\nn \\
\Tr(\rS^7) & = & -\frac{1}{64} \dimg \left(\left(21\dimg+1\right)+4\left(\dimg-3\right)\left(17\dimg-279\right)\ha\hb\hg\right.\nn\\
&&\left.+4\left(\dimg-3\right)\left(14\dimg^2-407\dimg+108\right)\ha^2\hb^2\hg^2\right),\label{TrRC}\\
\Tr(\rS^8) & = & \frac{1}{64} \dimg \left(2\left(11\dimg-1\right)+\left(\dimg-3\right)\left(91\dimg-2003\right)\ha\hb\hg+\right.\nn\\
&&\left.2\left(\dimg-3\right)\left(60\dimg^2-2430\dimg+1549\right)\ha^2\hb^2\hg^2+\right.\nn\\
&&\left.12\left(\dimg-3\right)\left(\dimg-1\right)^2\left(4\dimg-231\right)\ha^3\hb^3\hg^3\right),\nn\\
\Tr(\rS^9) & = & -\frac{1}{256} \dimg \left(\left(85\dimg+1\right)+400\left(\dimg-3\right)\left(\dimg-29\right)\ha\hb\hg+\right.\nn\\
&&\left.2\left(\dimg-3\right)\left(312\dimg^2-16354\dimg+5689\right)\ha^2\hb^2\hg^2+\right.\nn\\
&&\left.4\left(\dimg-1\right)\left(\dimg-3\right)\left(80\dimg^2-5792\dimg-303\right)\ha^3\hb^3\hg^3\right)\nn\\
\Tr(\rS^{10}) & = & \frac{1}{512} \dimg \left(4\left(43\dimg-1\right)+\left(\dimg-3\right)\left(938\dimg-37227\right)\ha\hb\hg\right.\nn\\
&&\left.+\left(\dimg-3\right)\left(1832\dimg^2-132662\dimg+56153\right)\ha^2\hb^2\hg^2+\right.\nn\\
&&\left.2\left(\dimg-3\right)\left(736\dimg^3-75674\dimg^2+96227\dimg-38215\right)\ha^3\hb^3\hg^3+\right.\nn\\
&&\left.24\left(\dimg-1\right)^3\left(\dimg-3\right)\left(16\dimg-2139\right)\ha^4\hb^4\hg^4\right)\nn
\eea

The dimensions of the irreps corresponding to the eigenvalues of the $\rS$ are obtained by calculating the respective projectors by \eqref{GenPr}, where
\begin{equation}
\begin{gathered}
A=\rS,\ \ I=\mathbb{S}_{[2,1]'},\ \ p=11,\\
 a_1=-\frac{1}{2},\ \ a_2=-1,\ \ a_3=-\frac{1}{2}+\ha,\ \ a_4=-\frac{1}{2}+\hb,\ \ a_5=-\frac{1}{2}+\hg,\\ a_6=-\frac{1}{2}-\ha,\ \ a_7=-\frac{1}{2}-\hb,\ \ a_8=-\frac{1}{2}-\hg,\ \ a_9=-\frac{3}{2}\ha,\ \ a_{10}=-\frac{3}{2}\hb,\ \ a_{11}=-\frac{3}{2}\hg
\end{gathered}
\end{equation}
and using \eqref{TrRC}. The results read
\bea
\dim_{-\frac{1}{2}} & = 2 X_2  = &  2 \times \frac{1}{2}\,\dimg \left( \dimg -3\right), \nn\\
\dim_{-1} & = 2 X_1 = & 2 \times\, \dimg, \nn \\
\dim_{\hat{\alpha}-\frac{1}{2}} & =  B =& \frac{ (\ha-1)(\hb-1)(\hg-1)(2\,\ha+\hb)(2\,\ha+\hg)(2\, \hb+1)(2\, \hg+1)(3\, \hb-1)(3\,\hg-1)}{8\, \ha^2\,(\ha-\hb)(\ha-\hg)(2\, \hb-\hg)(2\hg-\hb)\,\hb^2\,\hg^2},\nn \\
\dim_{\hat{\beta}-\frac{1}{2}} & =  B' =& \frac{ (\hb-1)(\hg-1)(\ha-1)(2\,\hb+\hg)(2\,\hb+\ha)(2\, \hg+1)(2\, \ha+1)(3\, \hg-1)(3\,\ha-1)}{8\, \hb^2\,(\hb-\hg)(\hb-\ha)(2\, \hg-\ha)(2\ha-\hg)\,\hg^2\,\ha^2},\nn  \\
\dim_{\hat{\gamma}-\frac{1}{2}} & = B'' =& \frac{ (\hg-1)(\ha-1)(\hb-1)(2\,\hg+\ha)(2\,\hg+\hb)(2\, \ha+1)(2\, \hb+1)(3\, \ha-1)(3\,\hb-1)}{8\, \hg^2\,(\hg-\ha)(\hg-\hb)(2\, \ha-\hb)(2\hb-\ha)\,\ha^2\,\hb^2},\nn \\
\dim_{-\hat{\alpha}-\frac{1}{2}} & =  Y_2 =& -\frac{(3\ha-1)(\hb-1)(\hg-1)(2\hb+1)(2\hg+1)}{
8 \ha^2 (\ha-\hb)(\ha-\hg)\hb\hg}, \label{uni03} \\
\dim_{-\hat{\beta}-\frac{1}{2}} & =  Y_2' =& -\frac{(3\hb-1)(\hg-1)(\ha-1)(2\hg+1)(2\ha+1)}{
	8 \hb^2 (\hb-\hg)(\hb-\ha)\hg\ha}, \nn \\
\dim_{-\hat{\gamma}-\frac{1}{2}} & =  Y_2'' =& -\frac{(3\hg-1)(\ha-1)(\hb-1)(2\ha+1)(2\hb+1)}{
	8 \hg^2 (\hg-\ha)(\hg-\hb)\ha\hb},\nn \\
\dim_{- \frac{3}{2} \ha} & =  C = & -\frac{2}{3}\frac{(1+2\ha)(1+2\hb)(1+2\hg)(1-\hb)(1-\hg)(\hb+\hg)(2\hb+\hg)(2\hg+\hb)}{\ha^3\hb\hg(\ha-2\hb)(\ha-2\hg)(\ha-\hb)(\ha-\hg)}  \nn \\
\dim_{- \frac{3}{2} \hb} & =  C' = & -\frac{2}{3}\frac{(1+2\hb)(1+2\hg)(1+2\ha)(1-\hg)(1-\ha)(\hg+\ha)(2\hg+\ha)(2\ha+\hg)}{\hb^3\hg\ha(\hb-2\hg)(\hb-2\ha)(\hb-\hg)(\hb-\ha)} \nn \\
\dim_{- \frac{3}{2} \hg} & =  C'' = & -\frac{2}{3}\frac{(1+2\hg)
(1+2\ha)(1+2\hb)(1-\ha)(1-\hb)(\ha+\hb)
(2\ha+\hb)(2\hb+\ha)}{\hg^3\ha\hb(\hg-2\ha)
(\hg-2\hb)(\hg-\ha)(\hg-\hb)}.
\nn
\eea
The full dimension of  all representations is
\be
\frac{1}{3} \dimg \left(\dimg-1\right) \left(\dimg+1\right),
\ee
as it should be.

Thus,
\be
[21] {\cal L}=2 X_1\oplus 2 X_2 \oplus Y_2\oplus Y_2'\oplus Y_2'' \oplus B \oplus B'\oplus B'' \oplus  C \oplus C' \oplus C'' .
\ee

\subsection{The cases with reduced characteristic identity }
\begin{itemize}
	\item $\soa_n$ and $\spa_{2r}$ - for these algebras $B$ and $C''=0$ and, therefore, the characteristic identity reads
\begin{align*}
&\left( \rS+\frac{1}{2} \right) \left(\rS+1\right)\left( \rS +\frac{n-6}{2(n-2)}\right)\left( \rS +\frac{1}{n-2}\right) \times \\ &\left(\rS+\frac{n-4}{2(n-2)}\right) \left( \rS +\frac{n+2}{2(n-2)}\right)\left( \rS +\frac{n-3}{n-2}\right) \left(\rS-\frac{3}{2(n-2)}\right)\left(\rS+\frac{3}{n-2}\right)\mathbb{S}_{[2,1]'} =0 
\end{align*}
and
\begin{align*}
&\left( \rS+\frac{1}{2} \right) \left(\rS+1\right)\left( \rS +\frac{r+3}{2(r+1)}\right)\left( \rS -\frac{1}{2(r+1)}\right) \times \\ &\left(\rS+\frac{r+2}{2(r+1)}\right) \left( \rS +\frac{r-1}{2(r+1)}\right)\left( \rS +\frac{2r+3}{2(r+1)}\right) \left(\rS+\frac{3}{4(r+1)}\right)\left(\rS-\frac{3}{2(r+1)}\right)\mathbb{S}_{[2,1]'} =0 
\end{align*}
respectively.
	\item $\sla_n$ - for these algebras $C''=0$, and, moreover, $\left(\rS+\frac{1}{2}+\ha\right)=\left(\rS+\frac{1}{2}-\hb\right)=\left(\rS+\frac{n-2}{2n}\right)$, $\left(\rS+\frac{1}{2}+\hb\right)=\left(\rS+\frac{1}{2}-\ha\right)=\left(\rS+\frac{n+2}{2n}\right)$ and $\left(\rS+\frac{1}{2}+\hg\right)=\left(\rS+1\right)$. Hence, the characteristic identity here is
\begin{align*}
&\rS\left( \rS+\frac{1}{2} \right) \left(\rS+1\right)\left(\rS+\frac{n+2}{2n}\right) \left( \rS +\frac{n-2}{2n}\right)\left(\rS-\frac{3}{2n}\right)\left(\rS+\frac{3}{2n}\right)\mathbb{S}_{[2,1]'} =0 .
\end{align*}
	\item for the exceptional algebras $Y_2''=B=B'=0$. Moreover, $\left(\rS+\frac{1}{2}\right)=\left(\rS+\frac{3}{2}\hg\right)$, and one of these multipliers can be omitted. Thus, the characteristic identity id simplified to be:
\begin{equation*}
\left( \rS+\frac{1}{2} \right) \left(\rS+1\right)\left( \rS +\frac{1}{6}\right) \left(\rS+\frac{1}{2}+\hat{\alpha}\right) \left( \rS +\frac{2}{3} - \hat{\alpha}\right) \left(\rS+\frac{3}{2} \hat{\alpha}\right)\left(\rS+\frac{1}{4} - \frac{3}{2} \hat{\alpha}\right)\mathbb{S}_{[2,1]'} =0 .
\end{equation*}
\end{itemize}

\setcounter{equation}0
\section{Vogel decomposition\label{Vdeco}}
Due to Vogel, in the decomposition $ad^{\otimes 3}$ the following irreps appear
\bea
Symmetric & = & 2 X_1 \oplus X_2\oplus B\oplus B'\oplus B''\oplus Y_3\oplus Y_3'\oplus Y_3'' \\
Hook & = & 2 X_1 \oplus 2 X_2 \oplus Y_2\oplus Y_2'\oplus Y_2'' \oplus B\oplus B'\oplus B''\oplus C\oplus C'\oplus C'' \\
Anti-Symmetric & =& X_0 \oplus X_2 \oplus Y_2\oplus Y_2'\oplus Y_2''\oplus X_3\oplus X_3'\oplus X_3'' .
\eea

The dimensions of these subrepresentations are given by the universal formulas
(\ref{uni01}), (\ref{uni02}), (\ref{uni03}),
and for simple Lie algebras  read:\\
\begin{tabular}{|l||l||l|}
	\hline \hline
	Irreps/Algebra & $\mathfrak{\sla}_n$ &  $\mathfrak{so}_n$  \\
	\hline
	$X_0$ & 1& 1 \\
	$X_1$ & $n^2-1$ & $\frac{1}{2} n (n-1)$ \\
	$X_2$ &$\frac{1}{2}(n^2-1)(n^2-4)$&$ \frac{1}{8}(n+2)n(n-1)(n-3)$\\
	$Y_2$ & $\frac{1}{4} (n+3) n^2 (n-1) $ & $ \frac{1}{12}(n+2)(n+1)n(n-3)$  \\
	$Y'_2$ & $\frac{1}{4} (n+1) n^2 (n-3) $ & $ \frac{1}{24}n(n-1)(n-2)(n-3)$  \\
	$Y''_2$ & $n^2 -1 $ & $ \frac{1}{2}(n-1)(n+2)$ \\
	$B$ & $\frac{1}{4}(n+1) n^2 (n-3)$ &  0  \\
	$B'$ & $\frac{1}{4}(n+3)n^2(n-1)$& $\frac{1}{8}(n+4)(n^2-1)(n-2)$ \\
	$B''$ & $\frac{1}{9}(n^2-1)^2 (n^2-9)$ & $\frac{1}{80} n(n^2-1)(n^2-4)(n-5)$ \\
	$C$ & $\frac{1}{9} (n+4)n^2 (n-1)(n^2-4)$ & $\frac{1}{45} n^2 (n^2-4)(n^2-16)$  \\
	$C'$ & $\frac{1}{9} (n+1) n^2 (n^2-4)(n-4)$ & $\frac{1}{144} n (n-1)(n^2-4)(n-3)(n-5)$ \\
	$C''$ & 0 & 0  \\
	$Y_3$ & $\frac{1}{36} (n+5)(n+1)^2 n^2 (n-1) $ & $\frac{1}{144} (n+4)(n+3)(n+2)(n-1)(n-2)(n-3)$  \\
	$Y'_3$ & $\frac{1}{36}(n+1)n^2 (n-1)^2(n-5) $ & $\frac{1}{720}n(n-1)(n-2)(n-3)(n-4)(n-5)$ \\
	$Y''_3$ & 1 & 0  \\
	$X_3$ & $\frac{1}{9}(n^2-1)^2(n^2-9) $ &$\frac{1}{144}n^2(n^2-1)(n^2-3n -10)$ \\
	$X'_3$ & $\frac{1}{18}(n^2-1)(n^2-4)(n^2-9) $ &$\frac{1}{72} n (n^2-1)(n-3)(n^2-16)$ \\
	$X''_3$ & 0 &0\\
	\hline
\end{tabular}
\\[0.5cm]
\begin{center}
\begin{tabular}{|l||l||l||l||l||l|}
	\hline \hline
	Irreps/Algebra & $\mathfrak{g}_2$ & $\mathfrak{f}_4$ & $\mathfrak{e}_6$ & $\mathfrak{e}_7$ & $\mathfrak{e}_8$ \\
	\hline
	$X_0$ &1 & 1 & 1 & 1 & 1\\
	$X_1$ & 14 & 52 & 78 & 133 & 248 \\
	$X_2$ & 77 & 1274 & 2925 & 8645 & 30380 \\
	$Y_2$ & 77 & 1053 & 2430 & 7371 & 27000 \\
	$Y'_2$ & 27 & 324 & 650 & 1539 & 3875 \\
	$Y''_2$ & 0 & 0 & 0 & 0 & 0 \\
	$B$ & 0 & 0 & 0 & 0 & 0 \\
	$B'$ & 0 & 0 & 0 & 0 & 0 \\
	$B''$ & 189 & 10829 & 34749 & 152152 & 779247 \\
	$C$ & 448 & 29172 & 105600 & 573440 & 4096000 \\
	$C'$ & 64 & 4096 & 11648 & 40755 & 147250 \\
	$C''$ & -77 & -1274 & -2925 & -8645 & -30380 \\
	$Y_3$ & 273 & 12376 & 43758 & 238602 & 1763125 \\
	$Y'_3$ & 7 & 273 & 650 & 1463 & 0 \\
	$Y''_3$ & -14 & -52 & -78 & -133 & -248 \\
	$X_3+X'_3+X''_3$ & 182 & 19448 & 70070 & 365750 & 2450240\\
	\hline
\end{tabular}
\\[0.5cm]
\end{center}
For the algebras $\spa_{2r}$, these dimensions can be acquired from those of $\soa_n$ by substituting $n\mapsto -2r$.

Note that for the exceptional algebras $\mathfrak{g}=\mathfrak{g}_2,\ \mathfrak{f}_4,\ \mathfrak{e}_6,\ \mathfrak{e}_7,\ \mathfrak{e}_8$ the dimensions of the representations $X_3$, $X_3'$ and $X_3''$ are not written in the table separately. Although $\dim X_3'$ and $\dim X_3''$ can be calculated to equal zero; computation of $\dim X_3$ yields an expression of the form $0/0$. The sum of the universal formulas for the dimensions of $X_3$, $X_3'$ and $X_3''$ is well-defined and presented in the second table.

Also, observe that $\dim Y_3''=-\dim X_1$ and $\dim C''=-\dim X_2$ for exceptional algebras are negative. This is another indication of the fact that the polynomials on the left hand side of \eqref{sS} and \eqref{rS} in this case are not minimal for $\sS$ and $\rS$ respectively, and the representations $X_1$ and $Y_3''$, as well as $X_2$ and $C''$, cannot be extracted separately.

\section{Acknowledgments}

The authors would like to thank R.L. Mkrtchyan and M.Y. Avetisyan for stimulating discussions. S.O.K. acknowledges the support of the Russian
Foundation for Basic Research, grant No. 20-52-12003.

\setcounter{equation}0
\section*{Appendix A: Formulas for traces}
\def\theequation{A.\arabic{equation}}

 \newtheorem{pro2b}[pro1]{Proposition} 
\begin{pro2b}\label{pro2b}
For all simple Lie algebras the
operators (\ref{03}), (\ref{04}) satisfy the identities:
\be\label{06i}
\begin{array}{c}
\Tr_{123} \hC_{[3]}  =
 \frac{1}{2} \dimg\; (\dimg+2) \; ,
 \;\;\;\;
\Tr_{123} \hC_{[1^3]} =
 \frac{1}{2} \dimg\; (2- \dimg) \; ,
 \end{array}
 \ee
 \be\label{06ii}
\Tr_{123} \; \mathbb{S}_{[1^3]}
 \widehat{C}_{12}^k =
 \frac{1}{3} (-\frac{1}{2})^k (\dimg-2)
 \dimg\; ,
 \ee
 \be\label{06iii}
  \Tr_{123} \; \mathbb{S}_{[3]}
 \widehat{C}_{12}^k =
 \frac{(-1)^k}{3} (\dimg+2)\,
 \Bigl( 1  +
  (\frac{\alpha}{2t})^k  \dim V_{(-\frac{\alpha}{2t})} +  (\frac{\beta}{2t})^k \dim V_{(-\frac{\beta}{2t})}
  + (\frac{\gamma}{2t})^k \dim V_{(-\frac{\gamma}{2t})}
  \Bigr)\; .
 \ee
 \end{pro2b}
 {\bf Proof.} Identities (\ref{06i}) are obtained
 by direct calculations with the use of the statements
 of Proposition {\bf \ref{pro2}} and relations
 (\ref{iskri01}).
 Identities (\ref{06ii}), (\ref{06iii})
 follow from a chain of
 equalities (here we use (\ref{chcp4}))
 $$
 \begin{array}{c}
  \Tr_{123} \; \mathbb{S}_{[1^3]}
 \widehat{C}_{12}^k = \frac{1}{6} (\dimg-2)
 \; \Tr_{12} \; (1 - P_{12})\widehat{C}_{12}^k=
 \frac{1}{3} (-\frac{1}{2})^k (\dimg-2)
 \dimg \; , \\ [0.2cm]
 \Tr_{123} \; \mathbb{S}_{[3]}
 \widehat{C}_{12}^k = \frac{1}{6} (\dimg+2)
 \; \Tr_{12} \; (1 + P_{12})\widehat{C}_{12}^k  = \\ [0.2cm] =
 \frac{1}{3} (\dimg+2)\,
 \Tr_{12}\widehat{C}_{12}^k \Bigl(
 P^{(+)}_{(-1)} + P^{(+)}_{(-\frac{\alpha}{2t})}
 + P^{(+)}_{(-\frac{\beta}{2t})} + P^{(+)}_{(-\frac{\gamma}{2t})}
 \Bigr) = \\ [0.2cm]
 = \frac{1}{3} (\dimg+2)\,
 \Bigl( (-1)^k  +
  (-\frac{\alpha}{2t})^k  \dim V_{(-\frac{\alpha}{2t})} +  (-\frac{\beta}{2t})^k \dim V_{(-\frac{\beta}{2t})}
  + (-\frac{\gamma}{2t})^k \dim V_{(-\frac{\gamma}{2t})}
  \Bigr)\; .
 \end{array}
 $$
 where we take into account (\ref{trac1}),
 (\ref{dim02a}), (\ref{dim02b}), (\ref{dim02c})
 \hfill \qed
 \vspace{0.2cm}

 For special cases of (\ref{06iii}), we obtain
 \be
 \lb{C12}
  \Tr_{123} \; \mathbb{S}_{[3]}
 \widehat{C}_{12} = \frac{1}{6} (\dimg+2)
 \dimg
 \; , \;\;\;
 \Tr_{123} \; \mathbb{S}_{[3]}
 \widehat{C}_{12}^2 = \frac{1}{4} (\dimg+2)
 \dimg \; ,
 \ee
   \be
 \lb{C34}
 \begin{array}{c}
   \Tr_{123} \; \mathbb{S}_{[3]}
 \widehat{C}_{12}^3 = \frac{1}{3} (\dimg+2)
 \Tr_{12}\widehat{C}_{+}^3 =
 - \frac{1}{24} (\dimg+2) \dimg \; ,
 \\ [0.2cm]
 \Tr_{123} \; \mathbb{S}_{[3]}
 \widehat{C}_{12}^4 = \frac{1}{3} (\dimg+2)
 \Tr_{12}\widehat{C}_{+}^4 =
 \frac{1}{48} (\dimg+2) \dimg
 \Bigl(7+ \frac{3 \alpha\beta\gamma}{2t^3}
 (\dimg-3) \Bigr) \; ,  \\ [0.2cm]
 \Tr_{123} \; \mathbb{S}_{[3]}
 \widehat{C}_{12}^5 = \frac{1}{3} (\dimg+2)
 \Tr_{12}\widehat{C}_{+}^5 =
 - \frac{1}{96} (\dimg+2) \dimg
 \Bigl(9+ \frac{2 \alpha\beta\gamma}{t^3}
 (\dimg-3) \Bigr)
 \end{array}
 \ee


 Below we also need the following   universal identities
for the square of the 3-split Casimir operator
 $$
 \begin{array}{c}
 \Tr_{123} (\hC_{[3]})^2 = \Tr_{123} \mathbb{S}_{[3]}
 (\widehat{C}_{12} + \widehat{C}_{13}
+ \widehat{C}_{23})^2 =
\frac{1}{4} \dimg\; (3 \dimg + 16) \; , \\ [0.2cm]
\Tr_{123} (\hC_{[1^3]})^{2} = \Tr_{123} \mathbb{S}_{[1^3]}
 (\widehat{C}_{12} + \widehat{C}_{13}
+ \widehat{C}_{23})^2 =
 \frac{1}{4} \dim^2 \mathfrak{g} \; ,
\end{array}
$$
To deduce these identities,  we take into account the relations
$\Tr_{123} (\mathbb{S}_{[3]} \, \hC_{ij}) =
\Tr_{123} (P_{ik} \mathbb{S}_{[3]} \, P_{ik} \, \hC_{ij}) =
\Tr_{123} (\mathbb{S}_{[3]} \, \hC_{kj})$ etc.
Note that by using $\Tr_{k} \bigl(P_{ik}  \widehat{C}_{ik})
= C_{i}$ we have
$$
 \Tr_1 (C^2_1) =
\Tr_{123} \bigl(P_{12}  P_{23} \widehat{C}_{12}
\widehat{C}_{13} \bigr) = \Tr_{123} \bigl(P_{13} P_{12} \widehat{C}_{12}
\widehat{C}_{13} \bigr) = \Tr_{13} \bigl(P_{13}
\Tr_{2}(P_{12} \widehat{C}_{12}) \widehat{C}_{13} \bigr) =
\Tr_{13} \bigl(P_{13} \widehat{C}_{13} \bigr) =
\dimg \; .
$$

One can also analytically deduce
\be\label{06ai}
\begin{array}{c}
\Tr_{123} (\hC_{[3]})^3 = \Tr_{123} \; \mathbb{S}_{[3]}
 (\widehat{C}_{12} + \widehat{C}_{13}
+ \widehat{C}_{23})^3
= -\frac{1}{8} \dimg\; (\dimg-58)
\; ,
\\ [0.2cm]
\Tr_{123} (\hC_{[1^3]})^3 = \Tr_{123} \mathbb{S}_{[1^3]}
 (\widehat{C}_{12} + \widehat{C}_{13}
+ \widehat{C}_{23})^3 =
- \frac{1}{8} \dimg\; (\dimg+6) \; ,
 \end{array}
 \ee

\setcounter{equation}0
\def\theequation{B.\arabic{equation}}
\section*{Appendix B: $ad^{\otimes 3}$ for $\mathfrak{\sla}_n$}

The tensor product of 3 adjoint representations of
$\mathfrak{\sla}_n$ is decomposed as follows:
 \be
 \lb{prod3}
\begin{array}{c}
([2,1^{n-2}])^{\otimes 3} = \Bigl( [\emptyset]
+ 2 \cdot  [2,1^{n-2}] + [2^2,1^{n-4}] + [3,1^{n-3}]
 + [3^2,2^{n-3}] + [4,2^{n-2}] \Bigr)
 \otimes [2,1^{n-2}] = \\ [0.3cm]
 = \Bigl( 2 \, \bigl( [\emptyset] + [2^2,1^{n-4}] + [3,1^{n-3}]
 + [3^2,2^{n-3}] + [4,2^{n-2}]\bigr) + 5 \cdot  [2,1^{n-2}]
 + \\ [0.3cm]
 + [2^2,1^{n-4}]\otimes [2,1^{n-2}] +
 [3,1^{n-3}]\otimes [2,1^{n-2}]
 + [3^2,2^{n-3}]\otimes [2,1^{n-2}] +
 [4,2^{n-2}] \otimes [2,1^{n-2}]\Bigr) ,
 \end{array}
 \ee
 where the diagrams $[2,1^{n-2}]$ and $[\emptyset]$
 correspond to the adjoint  representation and singlet.
 We recall (see e.g. \cite{IsKri}) that  $[2,1^{n-2}]^{\otimes 2}$ is divided
 into antisymmetric and symmetric parts (for $n> 3$)
 $$
 \begin{array}{c}
 \wedge([2,1^{n-2}]^{\otimes 2}) =
   [2,1^{n-2}] \oplus  [3,1^{n-3}] \oplus [3^2,2^{n-3}]
 \; , \\ [0.3cm]
 S([2,1^{n-2}]^{\otimes 2}) = [\emptyset] \oplus
  [2,1^{n-2}] \oplus [4,2^{n-2}] \oplus
 [2^2,1^{n-4}] \; .
 \end{array}
 $$
 In the expansion (\ref{prod3}) we have only the diagrams
 with $2n$ and $n$ boxes. For the products in the
 last line in (\ref{prod3}) we obtain
 ($n>4$)
 $$
 \begin{array}{l}
 [2^2,1^{n-4}]\otimes [2,1^{n-2}]  = [4,3,2^{n-4},1] \;\; +
 \\ [0.2cm]
 + \; ([3^2,2^{n-3}] + [3,1^{n-3}])_{_{dual}} +
 2\cdot [2^2,1^{n-4}] + [2,1^{n-2}] +
 ([3^3,2^{n-5},1] + [3,2,1^{n-5}])_{_{dual}}
 + [2^{3},1^{n-6}] \\ [0.3cm]
 [3,1^{n-3}]\otimes [2,1^{n-2}]  =  \\ [0.3cm]
 = [5,2^{n-3},1] + [4,3,2^{n-4},1] + [4,2^{n-2}] +
 [4,1^{n-4}] + [3,2,1^{n-5}] +
 2\cdot [3,1^{n-3}] + [2^2,1^{n-4}]  + [2,1^{n-2}]  \\ [0.3cm]
 [3^2,2^{n-3}]\otimes [2,1^{n-2}] = {\rm dual \; to} \;\;
  [3,1^{n-3}]\otimes [2,1^{n-2}] \\ [0.3cm]
 [4,2^{n-2}] \otimes [2,1^{n-2}] = [6,3^{n-2}] +
 [5,2^{n-3},1] +  \\ [0.3cm]
 + [5,4,3^{n-3}] + [4,3,2^{n-4},1] +  2\cdot [4,2^{n-2}] +
  [3,1^{n-3}] + [3^2,2^{n-3}]  + [2,1^{n-2}]
 \end{array}
 $$
 Thus, the entire decomposition reads
 \begin{gather}
 ([2,1^{n-1}])^{\otimes 3}=2\cdot[\emptyset]+9\cdot[2,1^{n-1}]+6[2^2,1^{n-4}]+[2^3,1^{n-6}]+6\cdot[4,2^{n-2}]+4\cdot[4,3,2^{n-4},1]+[6,3^{n-2}]\nn\\
 6([3,1^{n-3}]+[3^2,2^{n-3}])_{_{dual}}+2([3^3,2^{n-5},1]+[3,2,1^{n-5}])_{_{dual}}+\\
 +([4,1^{n-4}]+[4^3,3^{n-4}])_{_{dual}}+2\cdot([5,2^{n-3},1]+[5,4,3^{n-3}])_{_{dual}}\nn
 \end{gather}
 The dimensions of the irreducible representations
 of $\mathfrak{\sla}_n$ listed here are as follows
 \begin{gather}\lb{dimyo}
 \dim \, [2,1^{n-2}] = n^2-1\;\;({\rm selfdual}), \;\;\; \dim \, [\emptyset] = 1\;\;({\rm selfdual}), \;\;\;
 \dim \, [2^2,1^{n-4}] = \frac{n^2(n+1)(n-3)}{4}  \;\;
 ({\rm selfdual}),\nn \\[0.0cm]
 \dim \, [3,1^{n-3}] =
 \dim \, [3^2,2^{n-3}] = \frac{(n^2-1)(n^2-4)}{4} \;\;
 ({\rm dual \; diagrams}), \;\;\;
 \dim \, [4,2^{n-2}] = \frac{n^2(n-1)(n+3)}{4} \; ({\rm selfdual})\nn\\[0.0cm]
 {\rm dim}([4,3,2^{n-4},1])  =
 \frac{1}{9}(n-1)^2(n+1)^2(n^2-9)
 \;\; ({\rm selfdual})  \; , \;\;\;
\nn \\[0.0cm]
  {\rm dim}([3^3,2^{n-5},1])  =  {\rm dim}([3,2,1^{n-5}])  =
  \frac{1}{18}n^2(n+1)(n^2-4)(n-4) \;\;
 ({\rm dual \; diagrams}) \; ,\\[0.0cm]
 {\rm dim}([2^{3},1^{n-6}]) =  \frac{1}{36}(n+1)n^2(n-1)^2(n-5)
 \;\; ({\rm selfdual}) \; , \;\;\;\nn \\[0.0cm]
 {\rm dim}([5,2^{n-3},1])=  {\rm dim}([5,4,3^{n-3}])=
 \frac{1}{18}n^2(n^2-4)(n+4)(n-1)
 \;\;({\rm dual\; diagrams}) , \;\;\;\nn   \\[0.0cm]
 {\rm dim}([4,1^{n-4}])={\rm dim}([4^3,3^{n-4})=\frac{1}{36}(n^2-4)(n^2-9)(n^2-1) \;\;({\rm dual\; diagrams})\;\; ,\nn
 \\[0.0cm]
 {\rm dim}([6,3^{n-2}])= \frac{1}{36} n^2(n+1)^2(n+5)(n-1)\;\;({\rm selfdual}) .\nn
  \end{gather}

 \vspace{0.2cm}

 The dimensions in (\ref{dimyo})  are in agreement with the
 table presented in Section {\bf \ref{Vdeco}} (see also \cite{MSV,Fog,Lan}). Indeed, a comparison of the dimensions \eqref{dimyo} with the table allows us to establish the correspondence between the representations in the universal decomposition and the representations denoted by the Young diagrams:
\begin{equation}
\begin{aligned}
&X_0=Y_3''=[\emptyset], & &X_1=Y_2''=[2,1^{n-2}], & &Y_2'=B=[2^2,1^{n-4}],\\
&Y_2=B'=[4,2^{n-2}],& &X_2=[3,1^{n-3}]\oplus [3^2,2^{n-3}], & &X_3=B''=[4,3,2^{n-4},1],\\
&C'=[3^3,2^{n-5},1]\oplus[3,2,1^{n-5}], & &Y_3'=[2^3,1^{n-6}], & &C=[5,2^{n-3},1]\oplus[5,4,3^{n-3}],\\
&X_3'=[4,1^{n-4}]\oplus[4^3,3^{n-4}],& & Y_3=[6,3^{n-2}].
\end{aligned}
\end{equation}

\section*{Appendix C: Higher Casimir operators in the representations $X_2$, $Y_2$, $Y_2'$, $Y_2''$}
\setcounter{equation}0
\def\theequation{C.\arabic{equation}}

In this section, we will obtain universal expressions for eigenvalues of higher Casimir operators in the representations $X_2$, $Y_2$, $Y_2'$, $Y_2''$. We use the approach which was developed in \cite{Okubo}.
The resulting formulas are analogous to those of \cite{MSV}, where such identities were obtained for the adjoint representation $\ad$.

Let $\{Y_A\}$ be a basis of the enveloping algebra $\mathcal{U}(\mathfrak{g})$ of a Lie algebra $\mathfrak{g}$. If the operator $\widetilde{C}=D^{AB}Y_A\otimes Y_B\in\mathcal{U}(\mathfrak{g})\otimes \mathcal{U}\mathfrak(g)$ is ad-invariant, then
\begin{equation}
	C=\Tr_2(\id \otimes \ad)\widetilde{C}=D^{AB}Y_A \Tr(Y_B)
\end{equation}
lies in the centre of $\mathcal{U}(\mathfrak{g})$. Here $\id$ is the identity operator on $\mathcal{U}(\mathfrak{g})$ and $\Tr_2$ denotes the trace over the second tensor factor in $\mathcal{U}(\mathfrak{g})\otimes \mathfrak{gl}_{V_{\ad}}$. Apparently, any power of the split Casimir operator $\widehat{C}$ (its components are given by $\hC^{a_1a_2}_{b_1b_2}=\mathsf{g}^{cd}\sum_c(X_c)^{a_1}{}_{b_1}(X_d)^{a_2}{}_{b_2}$, where $\mathsf{g}^{cd}$ is the inverse Killing metric) is ad-invariant, so substituting $\widehat{C}^k$ for $\widetilde{C}$, we can define higher Casimir operators as
\begin{equation}
	C_k:=\Tr_2(\id \otimes \ad)\widehat{C}^k=\sum_{\substack{c_1,\dots,c_k\\d_1,\dots,d_k}}\mathsf{g}^{c_1d_1}\dots\mathsf{g}^{c_kd_k}X_{c_1}\dots X_{c_k}\Tr\left(\ad(X_{d_1})\dots \ad(X_{d_k})\right).
\end{equation}
In what follows we are only interested in $C_k$ taken in the representations $T=X_2,\ Y_2,\ Y_2'$ and $Y_2''$:
\begin{equation}\label{highCasGen}
	T(C_k)=\Tr_2(T \otimes \ad)\widehat{C}^k=\sum_{\substack{c_1,\dots,c_k\\d_1,\dots,d_k}}\mathsf{g}^{c_1d_1}\dots\mathsf{g}^{c_kd_k}T(X_{c_1})\dots T(X_{c_k})\Tr\left(\ad(X_{d_1})\dots \ad(X_{d_k})\right).
\end{equation}
Assuming the irreducibility of these representations, we infer by the ad-invariance of $C_k$ and Schur's lemma that $T(C_k)$ acts as a scalar operator whose eigenvalue is denoted by $c_k^T$. Therefore, $T(C_k)=c_k^TI$, where $I$ is the identity operator acting on $V_T$. To write the forthcoming relations in a compact form, we introduce the generating function
\begin{equation}
	c^T(z)=\sum_{k=0}^{\infty}c^T_kz^k.
\end{equation}
In what follows we will find explicit expressions for $c^{X_2}(z)$ and $c^{Y_2}(z)$. Analogous formulas for $c^{Y_2'}(z)$ and $c^{Y_2''}(z)$ are obtained by permutations of Vogel's parameters.

\subsection*{The representation $X_2$}
In order to calculate \eqref{highCasGen} for the case $T=X_2$, we  make use of the fact that this representation takes part in the decomposition $\ad^{\otimes 2}=X_1\oplus X_2\oplus Y_2\oplus Y_2'\oplus Y_2''\oplus X_0$ of the tensor square of the adjoint representation for any simple complex Lie algebra. We  denote the corresponding projector from $V_{\ad}^{\otimes 2}$ to $V_{X_2}$ by $\projt_{X_2}:V_{\ad}^{\otimes 2}\to V_{\ad}^{\otimes 2}$. It is precisely the operator whose trace is given in \eqref{ad2mdim2}, for explicit formulas see, e.g., \cite{IsKri}. We write
\begin{equation}
	(X_2\otimes \ad)\hC=\big((\projt_{X_2}\circ\, \ad^{\otimes 2})\otimes \ad\big)\hC=(\projt_{X_2}\otimes \id)(\hC_{13}+\hC_{23})=(\projt_{X_2}\otimes \id)\hC_{(3)}-(\projt_{X_2}\otimes\id)\hC_{12}
\end{equation}
To rewrite the expression for $(X_2\otimes \ad)\hC$ in a more convenient form, we find the characteristic identity for $X_2\hC_{(3)}\equiv(\projt_{X_2}\otimes \id)\hC_{(3)}$:
\begin{gather}\label{X2Char}
	\left(X_2\hC_{(3)}+1\right)\left(X_2\hC_{(3)}+\frac{1}{2}\right)X_2\hC_{(3)}\left(X_2\hC_{(3)}+\frac{1}{2}+\hat{\alpha}\right) \left( X_2\hC_{(3)} +\frac{1}{2} + \hat{\beta}\right)\left( X_2\hC_{(3)} +\frac{1}{2} + \hat{\gamma}\right)\\
	\left(X_2\hC_{(3)}+\frac{1}{2}-\hat{\alpha}\right) \left( X_2\hC_{(3)} +\frac{1}{2} - \hat{\beta}\right)\left( X_2\hC_{(3)} +\frac{1}{2} - \hat{\gamma}\right)\left(X_2\hC_{(3)}+\frac{3}{2} \hat{\alpha}\right)\left(X_2\hC_{(3)}+\frac{3}{2}  \hat{\beta}\right)\left(X_2\hC_{(3)}+\frac{3}{2}  \hat{\gamma}\right)=0\nonumber
\end{gather}
The invariant projectors obtained from \eqref{X2Char} will be denoted by $\proj_{-1},\ \proj_{-\frac{1}{2}}$ etc. The lower index stands for the eigenvalue of $X_2\hC_{(3)}$ on the corresponding eigenspace. The identity \eqref{X2Char} corresponds to the decomposition
\begin{equation}
	(\projt_{X_2}\otimes \id)\circ\,\ad^{\otimes 3}=X_1\oplus 2X_2\oplus (X_3\oplus X_3'\oplus X_3'')\oplus Y_2\oplus Y_2'\oplus Y_2''\oplus B\oplus B'\oplus B''\oplus C\oplus C'\oplus C''.
\end{equation}
Expanding $X_2\hC_{(3)}$ in terms of projectors onto their eigenspaces, using (see \cite{IsKri}])
\begin{equation}
	(\projt_{X_2}\otimes\id)\hC_{12}=0,
\end{equation}
and the orthonormality of the projectors, we get the following expression for the $k$-th power of $(X_2\otimes \ad)\hC$:
\begin{align}\label{X2highCas1}
	(X_2\otimes \ad)\hC^k&=\left(-\frac{1}{2}\right)^k\proj_{-\frac{1}{2}}+\left(-1\right)^k\proj_{-1}+\left(-\frac{3}{2}\ha\right)^k\proj_{-\frac{3}{2}\ha}+\left(-\frac{3}{2}\hb\right)^k\proj_{-\frac{3}{2}\hb}+\left(-\frac{3}{2}\hg\right)^k\proj_{-\frac{3}{2}\hg}\nonumber\\
	&\left(-\frac{1}{2}-\ha\right)^k\proj_{-\frac{1}{2}-\ha}+\left(-\frac{1}{2}-\hb\right)^k\proj_{-\frac{1}{2}-\hb}+\left(-\frac{1}{2}-\hg\right)^k\proj_{-\frac{1}{2}-\hg}+\\
	&\left(-\frac{1}{2}+\ha\right)^k\proj_{-\frac{1}{2}+\ha}+\left(-\frac{1}{2}+\hb\right)^k\proj_{-\frac{1}{2}+\hb}+\left(-\frac{1}{2}+\hg\right)^k\proj_{-\frac{1}{2}+\hg}\nonumber
\end{align}
In order to proceed, we need to take the trace $\Tr_3$ of \eqref{X2highCas1}. First, we calculate the partial traces of the projectors. This task is easily done by noticing that since, for instance,
\begin{equation}
	\Tr_3\proj_{-1}=c \projt_{X_2}
\end{equation}
for some constant $c\in\mathbb{C}$, then
\begin{equation}
	\Tr_{123}\proj_{-1}=c\Tr_{12}\projt_{X_2},\qquad\implies\qquad c=\frac{\Tr\proj_{-1}}{\Tr_{12}\projt_{X_2}}.
\end{equation}
The universal expressions for $\Tr\proj_{-1}$ and $\Tr_{12}\projt_{X_2}$ can be found from \eqref{uniDimsSym} as $\dim_{-1}/2$ and $\dim_{-\frac{1}{2}}$ respectively. Performing analogous steps for every projector in \eqref{X2highCas1} yields
\begin{align}\label{X2partTr}
	\Tr_3\proj_{-\frac{1}{2}}&=2\projt_{X_2}\nonumber\\
	\Tr_3\proj_{-1}&=-\frac{\ha\hb\hg}{(1+\ha)(1+\hb)(1+\hg)}\projt_{X_2}\nonumber\\
	\Tr_3\proj_{-\frac{3}{2}\ha}&=\frac{32\hb\hg(\hb+\hg)(2\hb+\hg)(2\hg+\hb)}{3\ha(\ha-2)(\ha-\hb)(\ha-\hg)(\ha-2\hb)(\ha-2\hg)}\projt_{X_2}\nonumber\\
	\Tr_3\proj_{-\frac{3}{2}\hb}&=\frac{32\ha\hg(\ha+\hg)(2\ha+\hg)(2\hg+\ha)}{3\hb(\hb-2)(\hb-\ha)(\hb-\hg)(\hb-2\ha)(\hb-2\hg)}\projt_{X_2}\nonumber\\
	\Tr_3\proj_{-\frac{3}{2}\hg}&=\frac{32\hb\ha(\hb+\ha)(2\hb+\ha)(2\ha+\hb)}{3\hg(\hg-2)(\hg-\hb)(\hg-\ha)(\hg-2\hb)(\hg-2\ha)}\projt_{X_2}\nonumber\\
	\Tr_3\proj_{-\frac{1}{2}-\ha}&=\frac{(3\ha-2)\hb\hg}{(\ha+1)(\ha-2)(\ha-\hb)(\ha-\hg)}\projt_{X_2}\\
	\Tr_3\proj_{-\frac{1}{2}-\hb}&=\frac{(3\hb-2)\ha\hg}{(\hb+1)(\hb-2)(\hb-\ha)(\hb-\hg)}\projt_{X_2}\nonumber\\
	\Tr_3\proj_{-\frac{1}{2}-\hg}&=\frac{(3\hg-2)\hb\ha}{(\hg+1)(\hg-2)(\hg-\hb)(\hg-\ha)}\projt_{X_2}\nonumber\\
	\Tr_3\proj_{-\frac{1}{2}+\ha}&=\frac{(2\ha+\hb)(2\ha+\hg)(3\hb-2)(3\hg-2)}{(1+\ha)(\ha-\hb)(\ha-\hg)(2\hb-\hg)(2\hg-\hb)}\projt_{X_2}\nonumber\\
	\Tr_3\proj_{-\frac{1}{2}+\hb}&=\frac{(2\hb+\ha)(2\hb+\hg)(3\ha-2)(3\hg-2)}{(1+\hb)(\hb-\ha)(\hb-\hg)(2\ha-\hg)(2\hg-\ha)}\projt_{X_2}\nonumber\\
	\Tr_3\proj_{-\frac{1}{2}+\hg}&=\frac{(2\hg+\hb)(2\hg+\ha)(3\hb-2)(3\ha-2)}{(1+\hg)(\hg-\hb)(\hg-\ha)(2\hb-\ha)(2\ha-\hb)}\projt_{X_2}\nonumber
\end{align}
Finally, multiplying \eqref{X2highCas1} by $z^k$, summing over $k$ and taking the partial trace in view of \eqref{X2partTr}, we have
\begin{align*}
	&\sum_{k=0}^\infty c_k^{X_2}z^k\projt_{X_2}=\sum_{k=0}^{\infty}(X_2\otimes \ad)\hC^kz^k=\left(\frac{2}{2+z}-\frac{\ha\hb\hg}{(1+z)(1+\ha)(1+\hb)(1+\hg)}+\right.\\
	&\bigg[\frac{64\hb\hg(\hb+\hg)(2\hb+\hg)(2\hg+\hb)}{3\ha(2+3\ha z)(\ha-2)(\ha-\hb)(\ha-\hg)(\ha-2\hb)(\ha-2\hg)}+(\ha\leftrightarrow\hb)+(\ha\leftrightarrow\hg)\bigg]+\\
	&\bigg[\frac{2(3\ha-2)\hb\hg}{(2+(1+2\ha)z)(\ha+1)(\ha-2)(\ha-\hb)(\ha-\hg)}+(\ha\leftrightarrow\hb)+(\ha\leftrightarrow\hg)\bigg]+\\
	&\left.\Bigg[\frac{2(2\ha+\hb)(2\ha+\hg)(3\hb-2)(3\hg-2)}{(2+(1-2\ha)z)(1+\ha)(\ha-\hb)(\ha-\hg)(2\hb-\hg)(2\hg-\hb)}+(\ha\leftrightarrow\hb)+(\ha\leftrightarrow\hg)\Bigg]\right)\projt_{X_2},
\end{align*}
so
\begin{align*}
	c^{X_2}(z)&=\frac{2}{2+z}-\frac{\ha\hb\hg}{(1+z)(1+\ha)(1+\hb)(1+\hg)}+\\
	&\bigg[\frac{64\hb\hg(\hb+\hg)(2\hb+\hg)(2\hg+\hb)}{3\ha(2+3\ha z)(\ha-2)(\ha-\hb)(\ha-\hg)(\ha-2\hb)(\ha-2\hg)}+(\ha\leftrightarrow\hb)+(\ha\leftrightarrow\hg)\bigg]+\\
	&\bigg[\frac{2(3\ha-2)\hb\hg}{(2+(1+2\ha)z)(\ha+1)(\ha-2)(\ha-\hb)(\ha-\hg)}+(\ha\leftrightarrow\hb)+(\ha\leftrightarrow\hg)\bigg]+\\
	&\Bigg[\frac{2(2\ha+\hb)(2\ha+\hg)(3\hb-2)(3\hg-2)}{(2+(1-2\ha)z)(1+\ha)(\ha-\hb)(\ha-\hg)(2\hb-\hg)(2\hg-\hb)}+(\ha\leftrightarrow\hb)+(\ha\leftrightarrow\hg))\Bigg]
\end{align*}

\subsection*{The representation $Y_2$}
The projector from $\ad^{\otimes 2}$ to $Y_2$ in the decomposition $\ad^{\otimes 2}=X_1\oplus X_2\oplus Y_2\oplus Y_2'\oplus Y_2'\oplus X_0$ will be denoted by $\projt_{Y_2}:V_{\ad}^{\otimes 2}\to V_{\ad}^{\otimes 2}$. It is the operator whose trace is given in \eqref{ad2mdim1}, for explicit formulas see, e.g., \cite{IsKri}. We have
\begin{equation}
	(Y_2\otimes \ad)\hC=\big((\projt_{Y_2}\circ \ad^{\otimes 2})\otimes \ad\big)\hC=\projt_{Y_2}\otimes \id(\hC_{13}+\hC_{23})=(\projt_T\otimes \id)\hC_{(3)}-(\projt_{Y_2}\otimes\id)\hC_{12}.
\end{equation}
Analogously to the $X_2$-case, we calculate the characteristic identity for $Y_2\hC_{(3)}\equiv(\projt_T\otimes \id)\hC_{(3)}$:
\begin{equation}
\begin{gathered}\label{Y2Char}
	\left(Y_2\hC_{(3)}+\frac{1}{2}\right)\left(Y_2\hC_{(3)}+1\right)\left(Y_2\hC_{(3)}+\frac{1}{2}+\ha\right)\left(Y_2\hC_{(3)}+3\ha\right)\left(Y_2\hC_{(3)}+\frac{3}{2}\ha\right)\\
	\left(Y_2\hC_{(3)}+\frac{1}{2}-\hb\right)\left(Y_2\hC_{(3)}+\frac{1}{2}-\hg\right)=0.
\end{gathered}
\end{equation}
The projectors onto invariant subspaces of $Y_2\hC_{(3)}$ obtained from \eqref{Y2Char} will be denoted by $\proj_{-1},\ \proj_{-\frac{1}{2}},$ etc. The characteristic identity \eqref{Y2Char} corresponds to the decomposition
\begin{equation}
	(\projt_{Y_2}\otimes \id)\circ\, \ad^{\otimes 3}=X_1\oplus X_2\oplus Y_2\oplus Y_3\oplus C\oplus B'\oplus B''.
\end{equation}
Expanding $Y_2\hC_{(3)}$ in terms of the orthonormal projectors $\proj_{-1}$, $\proj_{-\frac{1}{2}},\dots$ and using (see, e.g., \cite{IsKri})
\begin{equation}
	(\projt_{Y_2}\otimes \id)\hC_{12}=-\ha\projt_{Y_2}=-\ha\left(\proj_{-\frac{1}{2}}+\proj_{-1}+\proj_{-\frac{1}{2}-\ha}+\proj_{-3\ha}+\proj_{-\frac{3}{2}\ha}+\proj_{-\frac{1}{2}+\hb}+\proj_{-\frac{1}{2}+\hg}\right)
\end{equation}
we have for the $k$-th power of $(Y_2\otimes \ad)\hC$:
\begin{equation}\label{Y2highCas1}
	\begin{aligned}
		(Y_2\otimes \ad)\hC^k&=\left(\ha-\frac{1}{2}\right)^k\proj_{-\frac{1}{2}}+\left(\ha-1\right)^k\proj_{-1}+\left(-\frac{1}{2}\right)^k\proj_{-\frac{1}{2}-\ha}+\left(-2\ha\right)^k\proj_{-3\ha}+\left(-\frac{1}{2}\ha\right)^k\proj_{-\frac{3}{2}\ha}\\
		&+\left(-\hg\right)^k\proj_{-\frac{1}{2}+\hb}+\left(-\hb\right)^k\proj_{-\frac{1}{2}+\hg},
	\end{aligned}
\end{equation}
In order to take the partial trace $\Tr_3$ of \eqref{Y2highCas1}, we first calculate the following auxiliary traces:
\begin{align}\label{Y2partTr}
	\Tr_3\proj_{-\frac{1}{2}}&=\frac{(\ha+1)(\ha-2)(\ha-\hb)(\ha-\hg)}{(3\ha-2)\hb\hg}\projt_{Y_2}\nonumber\\
	\Tr_3\proj_{-1}&=-\frac{\ha(\ha-2)(\ha-\hb)(\ha-\hg)}{(3\ha-2)(1+\hb)(1+\hg)}\projt_{Y_2}\nonumber\\
	\Tr_3\proj_{-\frac{1}{2}-\ha}&=\projt_{Y_2}\nonumber\\
	\Tr_3\proj_{-3\ha}&=\frac{(\ha-2)(5\ha-2)(2\hb+\hg)(2\hg+\hb)}{3\ha(3\ha-2)(2\ha-\hb)(2\ha-\hg)}\projt_{Y_2}\\
	\Tr_3\proj_{-\frac{3}{2}\ha}&=\frac{32(\ha+1)(\hb+\hg)(2\hb+\hg)(2\hg+\hb)}{3\ha(3\ha-2)(\ha-2\hb)(\ha-2\hg)}\projt_{Y_2}\nonumber\\
	\Tr_3\proj_{-\frac{1}{2}+\hb}&=-\frac{(\ha+1)(\ha-2)(\ha+2\hb)(\ha-\hg)(2\hb+\hg)(3\hg-2)}{\hb(\hb+1)(\ha-2\hg)(2\ha-\hg)(\hb-\hg)\hg}\projt_{Y_2}\nonumber\\
	\Tr_3\proj_{-\frac{1}{2}+\hg}&=-\frac{(\ha+1)(\ha-2)(\ha+2\hg)(\ha-\hb)(2\hg+\hb)(3\hb-2)}{\hg(\hg+1)(\ha-2\hb)(2\ha-\hb)(\hg-\hb)\hb}\projt_{Y_2}\nonumber
\end{align}
Finally, multiplying \eqref{Y2highCas1} by $z^k$, summing over $k$ and taking the partial trace in view of \eqref{Y2partTr}, we have
\begin{align*}
	&\sum_{k=0}^\infty c_k^{Y_2}z^k\projt_{Y_2}=\sum_{k=0}^{\infty}(Y_2\otimes \ad)\hC^kz^k=\left(\frac{2}{2+z}+\frac{2(\ha+1)(\ha-2)(\ha-\hb)(\ha-\hg)}{(2-(2\ha-1)z)(3\ha-2)\hb\hg}-\frac{\ha(\ha-2)(\ha-\hb)(\ha-\hg)}{(1-(\ha-1)z)(3\ha-2)(1+\hb)(1+\hg)}\right.\\
	&+\frac{(\ha-2)(5\ha-2)(2\hb+\hg)(2\hg+\hb)}{3(1+2\ha z)\ha(3\ha-2)(2\ha-\hb)(2\ha-\hg)}+\frac{64(\ha+1)(\hb+\hg)(2\hb+\hg)(2\hg+\hb)}{3\ha(2+\ha z)(3\ha-2)(\ha-2\hb)(\ha-2\hg)}\\
	&\left.-\frac{(\ha+1)(\ha-2)(\ha+2\hb)(\ha-\hg)(2\hb+\hg)(3\hg-2)}{\hb(1+\hg z)(\hb+1)(\ha-2\hg)(2\ha-\hg)(\hb-\hg)\hg}-\frac{(\ha+1)(\ha-2)(\ha+2\hg)(\ha-\hb)(2\hg+\hb)(3\hb-2)}{\hg(1+\hb z)(\hg+1)(\ha-2\hb)(2\ha-\hb)(\hg-\hb)\hb}\right)\projt_{Y_2}
\end{align*}
Thus,
\begin{align*}
	&c^{Y_2}(z)=\frac{2}{2+z}+\frac{2(\ha+1)(\ha-2)(\ha-\hb)(\ha-\hg)}{(2-(2\ha-1)z)(3\ha-2)\hb\hg}-\frac{\ha(\ha-2)(\ha-\hb)(\ha-\hg)}{(1-(\ha-1)z)(3\ha-2)(1+\hb)(1+\hg)}\\
	&+\frac{(\ha-2)(5\ha-2)(2\hb+\hg)(2\hg+\hb)}{3(1+2\ha z)\ha(3\ha-2)(2\ha-\hb)(2\ha-\hg)}+\frac{64(\ha+1)(\hb+\hg)(2\hb+\hg)(2\hg+\hb)}{3\ha(2+\ha z)(3\ha-2)(\ha-2\hb)(\ha-2\hg)}\\
	&-\frac{(\ha+1)(\ha-2)(\ha+2\hb)(\ha-\hg)(2\hb+\hg)(3\hg-2)}{\hb(1+\hg z)(\hb+1)(\ha-2\hg)(2\ha-\hg)(\hb-\hg)\hg}-\frac{(\ha+1)(\ha-2)(\ha+2\hg)(\ha-\hb)(2\hg+\hb)(3\hb-2)}{\hg(1+\hb z)(\hg+1)(\ha-2\hb)(2\ha-\hb)(\hg-\hb)\hb}
\end{align*}

\end{document}